\documentclass[a4paper,11pt]{article}
\usepackage{pos}
\usepackage[export]{adjustbox}
\usepackage{graphicx,wrapfig}
\graphicspath{{Figures}}

\title{Machine learning for  four-dimensional SU(3) lattice gauge theories}
\ShortTitle{Machine learning for 4$d$ SU(3) lattice gauge theories}

\author*[a]{Urs Wenger}

\affiliation[a]{Institute for Theoretical Physics\\
  Albert Einstein Center for Fundamental Physics\\
  University of Bern\\
  CH-3012 Bern\\
  Switzerland}

\emailAdd{wenger@itp.unibe.ch}

\abstract{In this review I summarize how machine learning can be used in lattice gauge
  theory simulations and what ap\-proaches are currently available to
  improve the sampling of gauge field configurations, with
  a focus on applications in four-dimensional SU(3) gauge
  theories. These include approaches based on generative
  machine-learning models such as (stochastic) normalizing flows and
  diffusion processes, and an approach based on renormalization group
  (RG) transformations, more specifically the machine learning of RG-improved gauge actions
  using gauge-equivariant convolutional neural networks.
  In particular, I present scaling results for a machine-learned
  fixed-point action in four-dimensional SU(3) gauge theory towards
  the continuum limit. The results include observables based on the
  classically perfect gradient-flow scales, which are free of
  tree-level lattice artefacts to all orders, and quantities related
  to the static potential and the deconfinement transition.}

\FullConference{The 42nd International Symposium on Lattice Field Theory (LATTICE2025)\\
2-8 November 2025\\
Tata Institute of Fundamental Research, Mumbai, India\\}

\begin{document}
\maketitle

\section{Introduction and motivation}
Machine learning is having an ever increasing impact on simulations of
lattice gauge theories. In these proceedings I review some of the machine-learning
applications with a focus on those enabling efficient generation of SU(3)
gauge field configurations in four spacetime dimensions. I will concentrate the review on those
generative machine-learning approaches which are new and either have the potential
for or have already demonstrated promising scaling towards useful
lattices.

In lattice gauge theories we typically consider partition functions of
the form
\begin{equation}
  Z({\beta})=\int{\cal D}U \exp\{-{\beta} A[U]\}
\end{equation}
where $A[U]$ is the gauge action depending on the SU($N$) gauge field $U$,
$\beta=2N/g^2$ is the (inverse) gauge coupling, and ${\cal D}U$ is the
Haar integration measure. Expectation
values for observables take the form
\begin{equation}
  \langle{\cal O}_{\xi}\rangle_\beta=\frac{1}{Z(\beta)}\int{\cal D}U \exp\{-\beta A[U]\} \, {\cal O}_{\xi}[U]
\end{equation}
where $\xi$ denotes the characteristic physical length scale associated with
the given observable $\cal O$. Expressed in units of the lattice
spacing $a$ the length scale $\xi$ becomes a dimensionless quantity $\xi/a$ which
diverges towards the continuum limit $a\rightarrow 0$.
\begin{figure}[b]
   \includegraphics[width=0.28\textwidth]{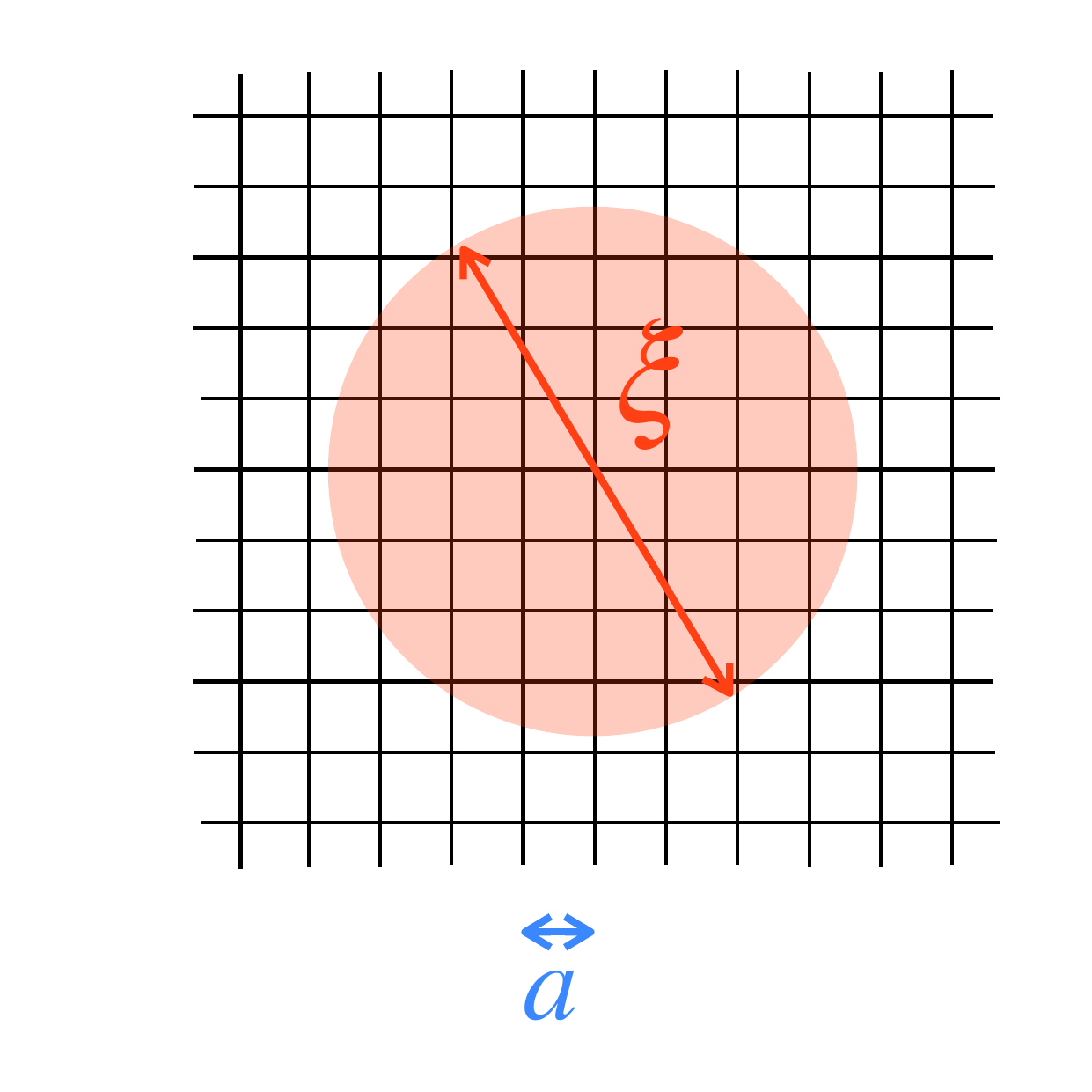}
  \hfill
 \includegraphics[width=0.28\textwidth]{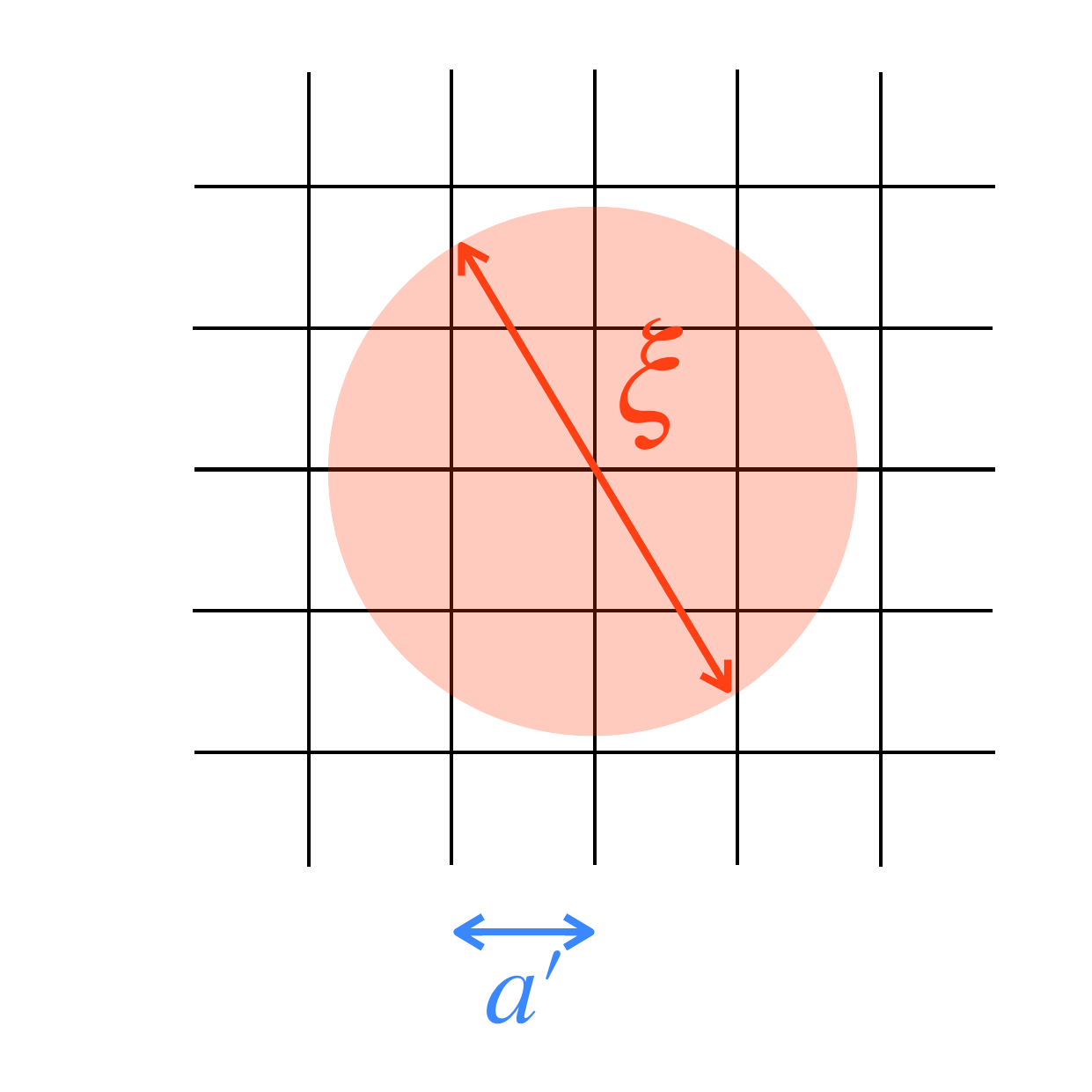}
  \hfill
  \includegraphics[width=0.28\textwidth]{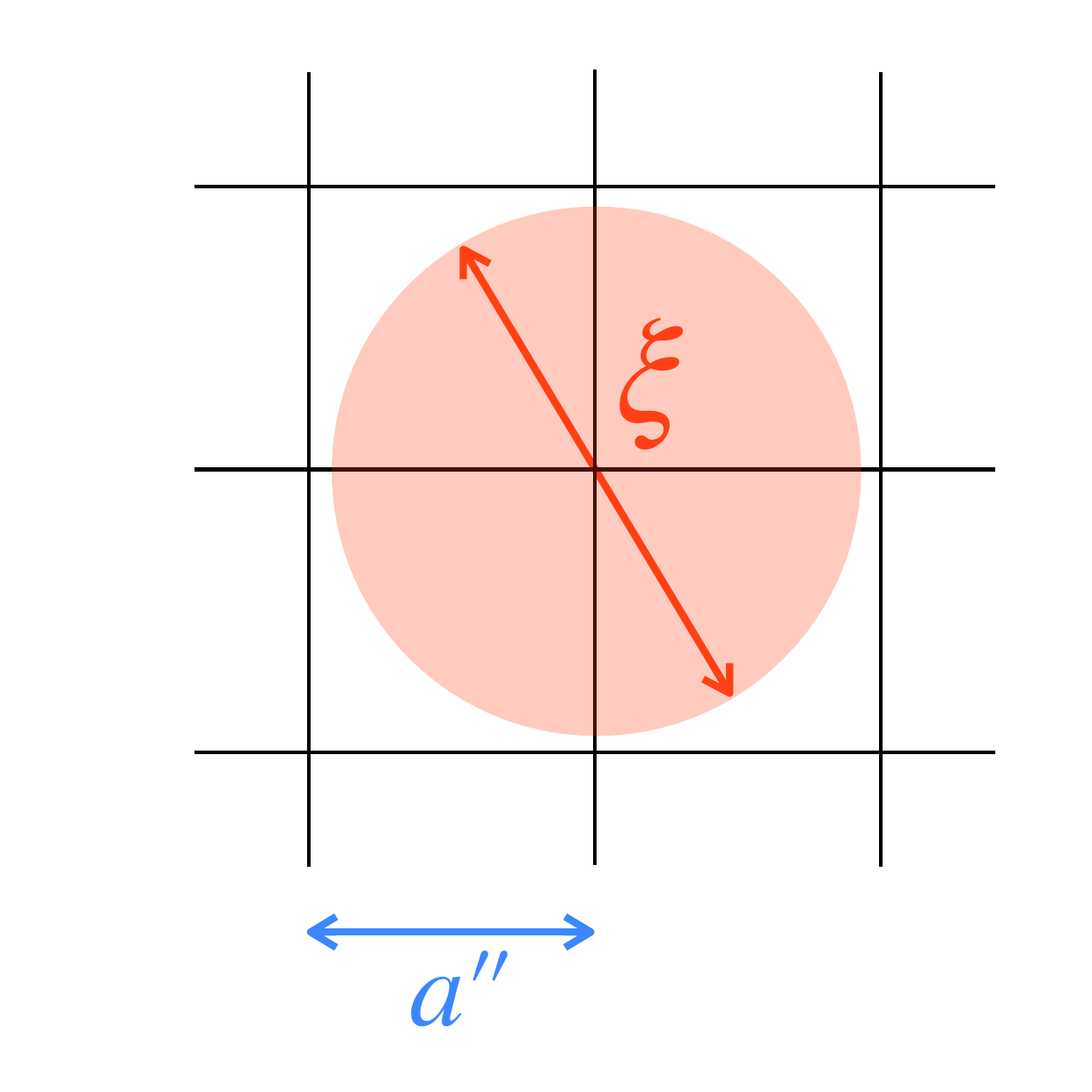}
  
  \caption{\label{fig:continuum limit}
    Illustration of the continuum limit for asymptotically free
    lattice field theories:
     the lattice spacing decreases from right to left, $a < a^{'} <
     a^{''}$, as the coupling decreases, $g <
     g^{'} < g^{''}$, or equivalently $\beta > \beta^{'} > \beta^{''}$ increases. In
    the limit  $\beta \rightarrow \infty$ the lattice spacing $a\rightarrow
0$ vanishes, which for a fixed physical length scale $\xi$ is equivalent to $\xi/a \rightarrow \infty$, i.e., the continuum limit is
realized as a second-order (continuous) phase transition.
  }
\end{figure}
The
lattice spacing is determined by the gauge coupling $\beta$ and for
asymptotically free field theories, such as four-dimensional SU($N$) gauge
theory, the continuum limit $a\rightarrow 0$ is achieved by taking
$\beta \rightarrow \infty$ or equivalently $g^2\rightarrow 0$, as illustrated in Fig.~\ref{fig:continuum limit}.
From the lattice perspective, the continuum limit  $\xi/a \rightarrow
\infty$ corresponds to a second-order (continuous) phase transition
where the physical correlation length $\xi$ diverges while the lattice
spacing is constant. As a consequence, simulations of lattice field
theories typically suffer from severe {\it critical slowing down}
towards the continuum limit resulting in a dramatic increase of
autocorrelation times. For lattice
gauge theories, critical slowing down may manifest itself in {\it
  topological freezing}. In such a situation, the simulations are stuck in sectors
of fixed topological charge, leading to meta-stabilities,
non-ergodicity of the simulations and eventually a failure to reach
the true equilibrium of the system.

This provides the motivation for applying machine-learning methods in lattice gauge
theories:
they try to avoid---in one way or
another---the critical slowing down encountered in Monte Carlo
simulations towards the continuum limit. Taking stock one can
identify
two complementary directions,  either to overcome critical
slowing down by employing generative machine-learning models at fine lattice
spacings, or to avoid critical slowing down
by simulating at coarse lattice spacing using a machine-learned  action
with highly suppressed lattice artefacts based on renormalization group transformations (RGTs): 
\begin{itemize}
\item Generative machine-learning models:\\
  These approaches try to generate uncorrelated gauge field
  configurations at fine lattice spacings. Currently there are two
  different procedures available based on mapping gauge field
  configurations from a prior distribution, for which it is easy
  to draw uncorrelated samples, to the
  target distribution. The machine-learned maps are based on 1) reversible
  normalising flows, or 2) backward diﬀusion processes. A new
  method 3) is based on
  non-equilibrium Markov Chain Monte Carlo (NE-MCMC) and  machine-learned stochastic normalising flows. I discuss these
  three approaches in some detail in in Secs.~\ref{sec:Normalizing
    flows}, \ref{sec:Diffusion models} and \ref{sec:Stochastic
    normalizing flows}, respectively.

\item Machine learning RGTs:\\
  In this approach one employs RGTs to relate fine to coarse
  lattices in order to construct effective RGT-improved lattice
  actions. Here, the idea is that uncorrelated lattice configurations can
  be generated at coarse lattice spacings, where there is no critical
  slowing down, while at the same time large lattice artefacts are
  avoided. Then the main challenge is to learn or parametrize the
  inverse RGT. I review this approach in Sec.~\ref{sec:Inverse
    renormalization group transformations}.
\end{itemize}

Apart from these {\it generative} machine-learning approaches, on which
this review is focused, there also exist machine-learning strategies
to improve the evaluation of observables. One strategy tries to
enhance the signal-to-noise ratios by employing, e.g., control
variates, surrogate variables, or contour deformations, and optimizing them using machine-learning. These strategies have been
reviewed and discussed in detail by Scott Lawrence in a plenary talk
at last year's lattice conference \cite{Lawrence:2025rnk}. 
An intriguing idea is to use the Feynman-Hellmann theorem and
employ operator insertions, derivative observables and machine-learned
normalizing flows which, in combination, can achieve a significant variance reduction \cite{Abbott:2026ylv}. Another interesting example is the
exploration of gauge-fixing schemes using machine learning
\cite{Detmold:2024mts}. Let me also point out the efforts to optimise
wave functions and ground-state operators. One example concerns the
parametrization of interpolators for a static quark-antiquark pair, in
order to optimize the overlap with the ground state \cite{Bellscheidt:2026rjh}. 
Another example is the determination of the energy
spectrum of SU($N$) gauge theories in the Hamiltonian formulation
using physics-informed neural networks
\cite{Spriggs:2025sea,Romiti:2025cfs}. The variety of these applications demonstrates
the versatility of machine learning as a tool to improve and enhance
simulations of lattice gauge theories.

\section{Generative machine-learning models}
\label{sec:generative ML models}
\subsection{Normalizing flows}
\label{sec:Normalizing flows}

The approach of using normalizing flows for generative machine-learning models, or more generally flow-based
samplers, has been reviewed by Gurtej Kanwar in his plenary talk at
the lattice conference in 2023 \cite{Kanwar:2024ujc}, including an extensive discussion
of the progress at the time and the prospects for the
future. The review provided here is therefore kept very brief, yet
it is interesting to understand what has happend in the meantime and
what the current status is.

The general idea is to learn a map from a prior distribution of
field configurations to the
target one, either using a continuous or a discrete flow of the
underlying field variables. More
formally, given a (simple) prior distribution $r(U)$ on the gauge fields $U$,
the (complicated) target distribution is $q(U') = q(f(U))$ where the
diffeomorphism $f(U)$ is found and described by a machine-learning
approach,
\begin{equation}
  r(U)  \,
  \stackrel{\text{ML}}{\xrightarrow{\hspace*{0.5cm}}}
  \, q(U') = q(f(U)) \, .
\end{equation}
Apart from finding suitable gauge-equivariant maps, the main problem lies in the
fact that the transformations require the computation of invertible
Jacobians $J_f$ and their determinants,
\begin{equation}
 q(f(U)) = r(U) \left| \det \frac{\partial f(U)}{\partial U}\right|^{-1}.
\end{equation}
The construction of calculationally tractable transformations is
usually achieved through a discrete set of coupling layers containing gauge-equivariant
functions $g_i$, such that
\begin{equation}
 f \equiv g_m \circ \cdots \circ g_1
\end{equation}
with $g_i$ being easily invertible. The Jacobian determinant of $f$
can then be calculated efficiently by
\begin{equation}
 \det J_f = \det J_{g_1} \cdot \ldots \cdot  \det J_{g_m}\,.
\end{equation}
This can be accomplished, for example, by transforming only a subset of
the degrees of freedom conditioned on the complementary
subset. Most effectively, a simple triangular Jacobian is obtained by a
suitable decoupling of the variables. The maps are self-trained by
minimizing a target loss function based on the reverse
Kullback-Leibler divergence
\begin{equation}
 D_{KL}(q||p) \equiv \int {\cal D}U \,q(U)\,[\log q(U) - \log p(U)]
\end{equation}
 which is available because the gauge field
action $S[U]$ and hence the exact target distribution $p(U)$ is
explicitly known. (Recall that $q(U)$ is the 
machine-learned model output
distribution and can be regarded as a variational parametrization
ansatz for the target distribution.)

The symmetries of the gauge action can
be taken into account by incorporating them into the flows. In
particular, it is
crucial to take gauge symmetry into account, such as in
Ref.~\cite{Kanwar:2020xzo, Boyda:2020hsi} where this approach has been
pioneered for U(1) and SU($N$) gauge theories in two spacetime
dimensions. For theories with dynamical fermions, flow-based sampling including
pseudofermions has also been studied~\cite{Abbott:2022zhs}.  In Ref.~\cite{Abbott:2023thq} the framework has been extended to SU($3$)
gauge theory in four spacetime dimensions with the latest
advancements reported in \cite{Abbott:2024mix,Abbott:2025kvi}.
New developments
include architectural progress and the use of the correlated
ensemble method \cite{Abbott:2024kfc}.

Reviewing the developments over the last couple of years it is
probably safe to
state that the scaling of the normalizing-flow approach in the context
of generative machine-learning models to four spacetime dimensions,
SU($N$) gauge theories, large volumes and fine lattice spacing all at
the same time has
turned out to be very challenging \cite{Abbott:2022zsh}. As a consequence, progress in this
direction has somewhat slowed down and more recent normalizing-flow applications
seem to be focusing on complementary directions \cite{Abbott:2024kfc,Abbott:2024mix,Abbott:2026ylv}.

\subsection{Diffusion models}
\label{sec:Diffusion models}
Generative diffusion models have become very popular in recent years and
have enabled many exciting applications in text-to-image and
text-to-video generation. For a discussion of the core principles I
refer to the review \cite{lai2025principlesdiffusionmodels} which also
discusses the relation of score- or energy-based diffusion models with the
flow-based models.

\begin{figure}[b]
  \includegraphics[angle=-90,width=\textwidth]{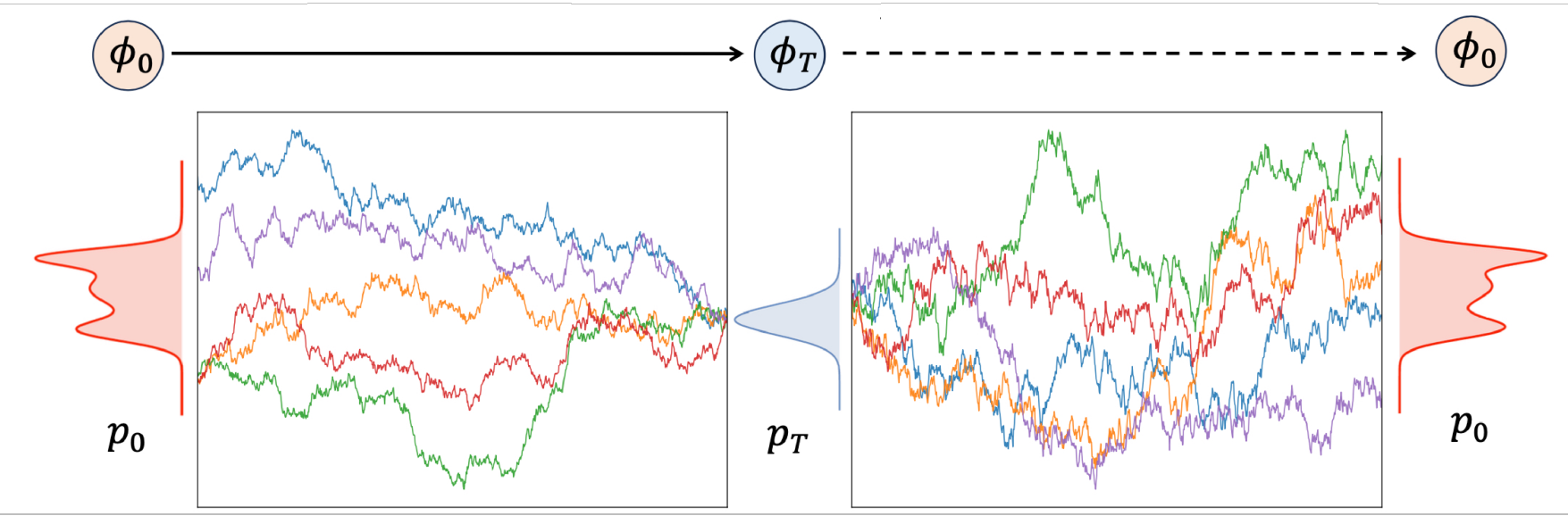}
  \caption{\label{fig:diffusion_process}
    Illustration of the forward diffusion process where
    stochastic noise is added and the
    backward denoising process employed in
    generative diffusion models. Figure taken from
Ref.~\cite{Wang:2023exq}.
  }
\end{figure}

In a nutshell, generative
diffusion models are based on
adding noise to the degrees of freedom $\phi_0$, e.g., a gauge field $\phi_0=U_0$
distributed according to a target distribution $p_0(U_0)$, using a
stochastic differential equation. The stochastic evolution of the
degrees of freedom in the forward process up to the fictitious time $T$ produces new
configurations $\phi_T = U_T$ distributed according to a simple,
easy-to-sample prior distribution $p_T(U_T)$. The forward process can be
reversed using a different, but related stochastic differential
equation mapping  $p_T$ back to $p_0$. This backward denoising process is used to
generate new samples $\phi_0 = U'$ distributed according to
$p_0$. The forward and backward processes are illustrated in
Fig.~\ref{fig:diffusion_process} taken from
Ref.~\cite{Wang:2023exq}. One of the challenges for diffusion-based
generative models is to guarantee exactness in the sense that the
generated samples are asymptotically distributed according to the
target distribution.   

To be more formal, the stochastic differential equation for the
forward process is
\begin{equation}
  \label{eq:forward process}
  \frac{d\phi}{dt} = f(\phi,t) + g(t) \eta(t)
\end{equation}
where $f(\phi,t)$ is a time-dependent drift term, $g(t)$ the time-dependent diffusion
coefficient, and $\eta(t)$ Gaussian noise. The denoising backward
process is described by the stochastic differential equation
\begin{equation}
    \label{eq:backward process}
  \frac{d\phi}{dt} = \left[ f(\phi,t) - g(t)^2 \nabla_\phi
    \log p_t(\phi) \right] + g(t) \,
  \bar \eta(t)
\end{equation}
where the drift term in square brackets now contains the gradient of
$\log p_t(\phi)$ defining the score function $\nabla_\phi
    \log p_t(\phi)$ to be learned with a machine-learning
    approach. 

One can gain an interesting insight if one sets the drift term
$f(\phi,t)=0$ in Eqs.~(\ref{eq:forward process}) and (\ref{eq:backward
  process}). In this case, the forward process corresponds to a so-called
variance-expanding scheme because the added noise is not regulated by
a drift term. On the other hand, the backward process becomes a variant
of stochastic quantization, see~Ref.~\cite{Fukushima:2024oij} for a pedagogical review. This relation
can then be used as a physical condition
for sampling configurations along the backward process, as
suggested in Ref.~\cite{Zhu:2025pmw}. There, such a physics-conditioned
diffusion model is discussed in the context of U$(1)$ gauge theory in
two spacetime dimensions combined with Metropolis-adjusted annealed
Langevin dynamics in order to guarantee the exactness of the sampling
and to enhance efficiency. Finally, we
note that the connection to stochastic quantisation has also been
exploited for complex-valued actions, using energy-based diffusions
models, see Ref.~\cite{Aarts:2025lpi}, although this is for
two-dimensional $\phi^4$ theory.

Another example of a diffusion-based machine-learning approach just
appeared before the conference \cite{Vega:2025hgz}. In this work, care
was taken to construct symmetry-preserving diffusion models based on
score matching. Moreover, the loss function is augmented with a regularization
term containing the force which, in general, is analytically known for
gauge field theories. Consequently, two avenues are suggested to make the
diffusion-based sampling exact, one based on reweighting and the other on
resampling.

While these diffusion-based approaches are very promising and show great
potential, they come with the caveat that so far they have been employed
in two-dimensional gauge theories only, mainly for U$(1)$.  Exceptions
are Ref.~\cite{lou2023scalingriemanniandiffusionmodels} for SU$(3)$
gauge theory on a $4^2$ lattice and the recent
work in Ref.~\cite{Kanwar:2025wuc}
for a single generic SU$(N)$ gauge matrix. Furthermore, the
extensions to two-dimensional U$(2)$ and SU$(2)$ gauge theories in
Ref.~\cite{Aarts:2026zzr,Alharazin:2026lcb} demonstrate that progress
is fast in this area. However, experience has shown that successful machine-learning approaches for
two-dimensional gauge theories do not easily transfer to SU$(3)$ and
four spacetime dimensions.

\subsection{Stochastic normalizing flows}
\label{sec:Stochastic normalizing flows}
\begin{figure}[t]
  \vspace*{-0.5cm}
  \centering
  \includegraphics[angle=-90,width=0.975\textwidth]{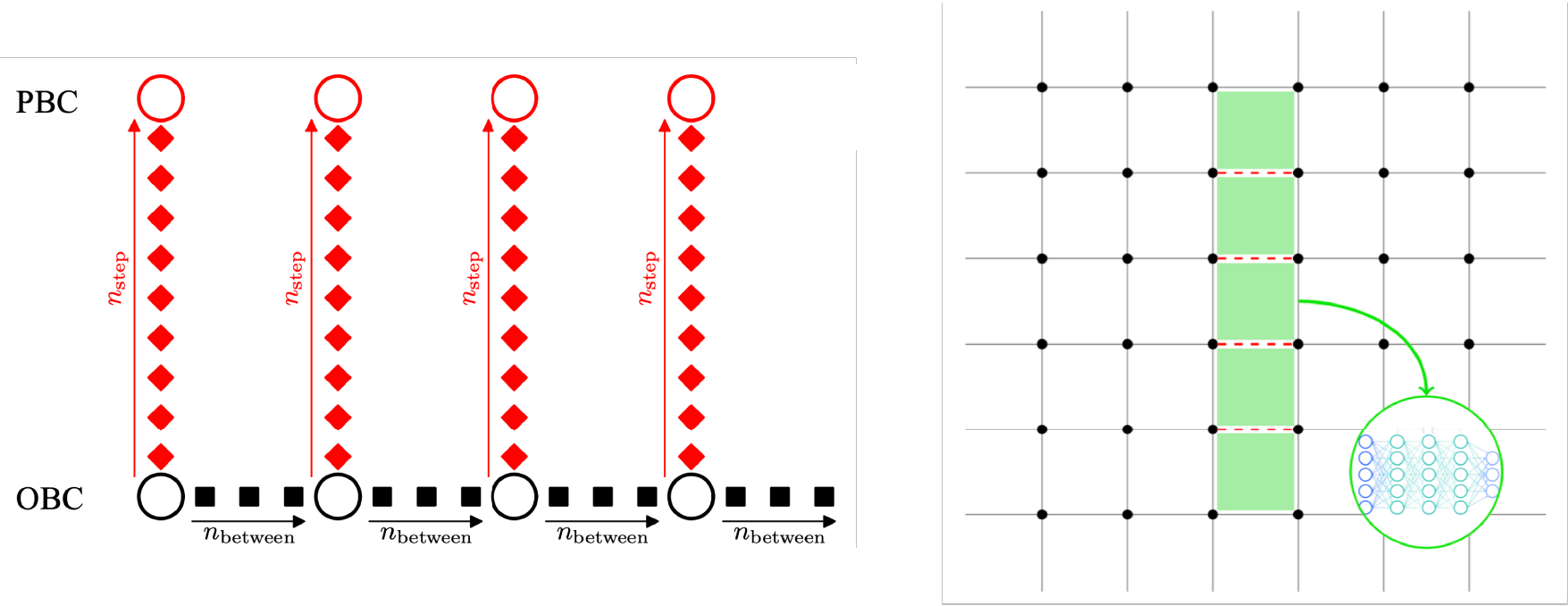}
  \caption{\label{fig:stochastic_flow_schedule}
    Illustration of the stochastic non-equilibrium MCMC scheme. {\it
      Left plot:}
    Configurations obtained on a lattice
    with open boundary conditions (OBC) using standard MCMC updates are
    connected via non-equilibrium MCMC updates to configurations on a lattice
    with periodic boundary conditions (PBC). {\it Right plot:}
    Connecting OBC to PBC on a localized
    defect using machine-learned stochastic normalizing flows on the links
    and plaquettes affected by the defect (plot taken from \cite{Bonanno:2026tle}).
  }
\end{figure}

A very interesting approach involving machine learning for lattice
gauge theory has been put forward in Ref.~\cite{Bonanno:2024udh} and is
applied in the context of four-dimensional SU$(3)$ gauge theory in Ref.~\cite{Vadacchino:2024lob}. The approach is based on
the fact that topological freezing can be mitigated by employing open
boundary conditions (OBC), e.g., in the direction of Euclidean time. Instead
of opening the boundaries on a full spatial timeslice, one can open it
on a spatially localized defect only. In order to avoid the
problems with the loss of translational invariance in time (and space
in case a spatially localized defect is used), one can apply parallel
tempering to connect the ensemble with the defect to the one with
fully periodic boundary conditions (PBC) \cite{Hasenbusch:2017unr,Bonanno:2020hht}. However, instead of using
expensive parallel tempering, one can employ out-of-equilibrium
evolutions based on Jarzynski’s equality
\cite{PhysRevLett.78.2690,Caselle:2016wsw}. This is illustrated in
the left plot of
Fig.~\ref{fig:stochastic_flow_schedule} where the black squares at the
bottom represent a number of Monte Carlo steps on the lattice
with OBC enabling an efficient update of topological modes.
After
every $n_\text{between}$ updates the
configuration so obtained on the
OBC lattice is subject to $n_\text{steps}$ of
out-of-equilibrium update steps (red squares) transforming it into a configuration
on the PBC lattice. So from an ensemble of topologically decorrelated OBC configurations one
obtains an ensemble of PBC configurations. The latter needs to be
reweighted with a statistical weight obtained from the work spent
along the non-equilibrium evolution. This evolution follows a specific
protocol with transition probabilities $P_i,
i=1,\ldots,N=n_\text{steps}$ satisfying detailed balance, but the
evolution does not
reach equilibrium.
Connecting the non-equilibrium Markov Chain Monte Carlo
(NE-MCMC) to
the theoretical framework of stochastic normalizing flows
\cite{Caselle:2022acb} eventually enables the application of machine-learning
techniques \cite{Bulgarelli:2024cqc,Bulgarelli:2024brv}. More
specifically, the NE-MCMC evolution can be enhanced by combining it with a class of
deep generative normalizing-flow models parametrized by $g_\theta$,
\begin{equation}
  U_0 \longrightarrow g_\theta^1(U_0) \stackrel{P_1}{\longrightarrow} U_1
  \longrightarrow g_\theta^2(U_1) \stackrel{P_2}{\longrightarrow} \ldots
  \stackrel{P_{N}}{\longrightarrow} U_{N} = U .
\end{equation}
This defines a particular instance of a stochastic normalizing flow
where the normalizing-flow layers are trained to bring the NE-MCMC
as close as possible to equilibrium, thereby reducing the variance of
the work done along the given NE-MCMC protocol and hence improving the
effectiveness of reweighting the configurations on the PBC
lattice. Applying the normalizing flow only on the gauge links
localized around the defect \cite{Bulgarelli:2024yrz}, as illustrated
in the right plot of Fig.~\ref{fig:stochastic_flow_schedule},
simplifies the ma\-chine-learning task considerably. By combining the
normalizing flow
with the NE-MCMC one can achieve a speed-up of about a factor $\sim 3$
compared to just the standard NE-MCMC \cite{Bonanno:2026tle}.

As demonstrated in Ref.~\cite{Bonanno:2025pdp} this combined
strategy scales well for four-dimensional SU(3) gauge theories, both
when decreasing the lattice spacing and increasing the number of
degrees of freedom. For example, one finds that scaling $n_\text{between}$ with
$a^{-2}$ keeps autocorrelations roughly fixed, while $n_\text{step}
\propto a^{-3}$ maintains the efficiency of the flow towards the
continuum limit. Indeed, scaling up to $\beta=6.4$ on a $34^4$ lattice
has been reported at this conference \cite{Bonanno:2026tle} which
underlines the potential of this approach.
\vspace*{-0.2cm}

\section{Inverse renormalization group transformations}
\label{sec:Inverse renormalization group transformations}
This machine-learning approach \cite{Holland:2023ews,Holland:2024muu} is based on the idea that at coarse
lattice spacings uncorrelated lattice configurations can be easily
generated without critical
slowing down, while at the same
time large lattice artefacts are avoided by machine learning a highly
improved gauge action based on the renormalization group.

\subsection{The fixed-point action}
\begin{figure}
  \vspace*{-0.5cm}
  \centering
  \includegraphics[width=0.7\textwidth]{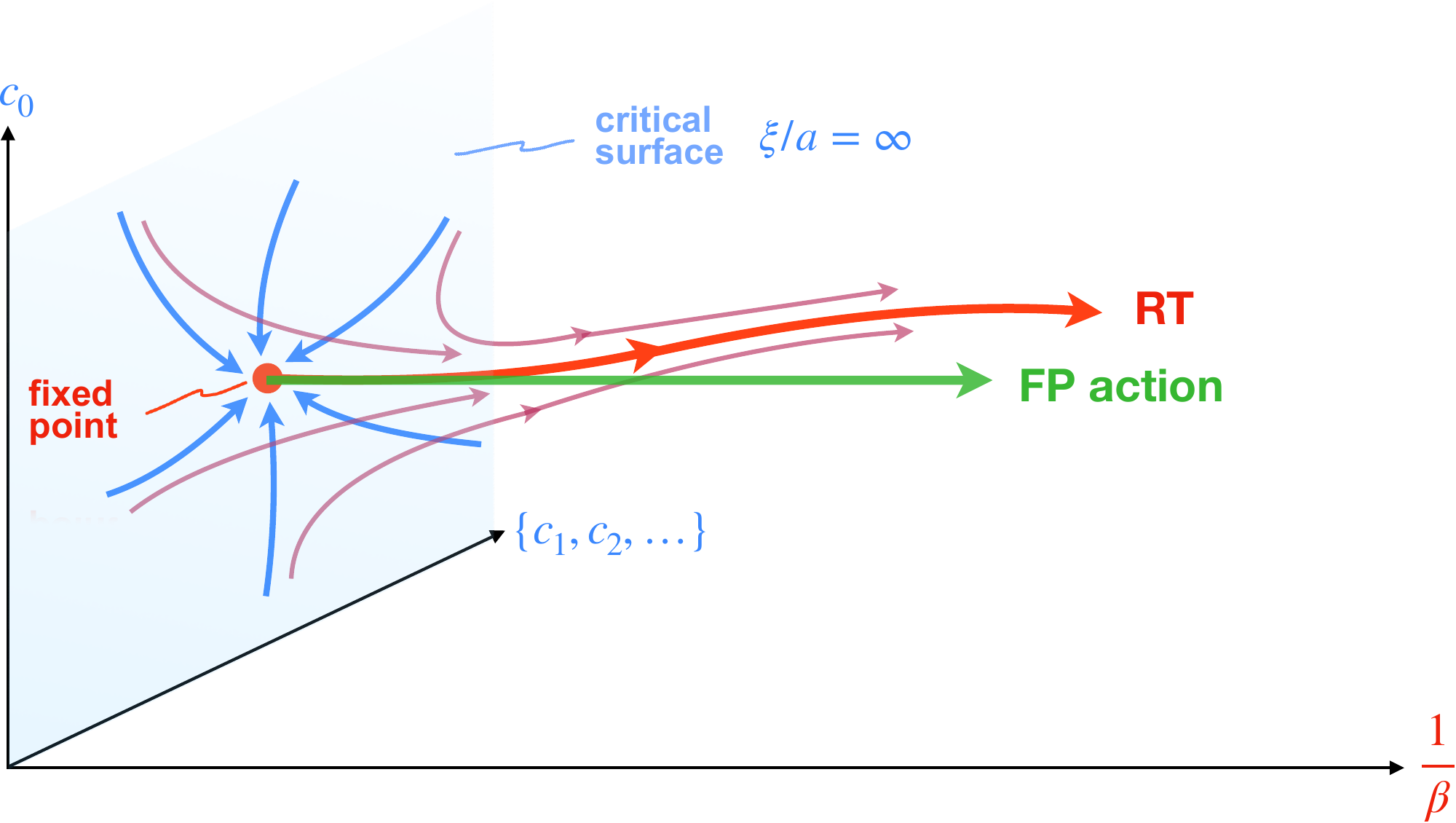}
  \caption{\label{fig:RGT flow}
    \
    Illustration of the RGT flow. The plane at $\beta \rightarrow
    \infty$ represents the critical surface where all the irrelevant
    couplings $\{c_\alpha\}$ flow to the fixed point (FP). An action
    perturbed from the FP in direction of the only
    relevant coupling $g^2 \sim 1/\beta$ flows, under repeated (continuous) RGTs, along the
    renormalized trajectory (RT) 
    defining {\it quantum perfect actions} with no lattice artefacts at finite lattice
    spacing. The action along the straight line in direction $1/\beta$ 
    emanating from the FP defines a {\it classically perfect action}
    with no tree-level lattice artefacts ${\cal O}(a^{2n})$ for all $n$. 
  }
\end{figure}

A real-space renormalization group transformation (RGT) can be defined
as 
\begin{equation}
  \exp\left\{-\beta' A'[V]\right\} = \int{\cal D}U \exp\left\{-\beta \left(A[U] + T[U,V]\right)\right\}
\end{equation}
where the blocking kernel $T[U,V]$ couples the gauge field $V$ on a
coarse lattice with lattice spacing $a'$ to the gauge field $U$ on a
fine lattice with lattice spacing $a$. This can be realized, for
example, by the kernel
\begin{equation}
  T[U,V] = - \frac{\kappa}{N} \sum_{x', \mu} \left\{ \text{Re} \text{Tr} \left(V_\mu(x') \cdot Q^\dagger_\mu(x') \right) - {\cal N}_\mu^\beta\right\}
\end{equation}
where $x'$ are the coordinates of
the coarse lattice, $Q_\mu(x')$ is a blocked
link constructed from the fine links $U(x)$, and ${\cal N}_\mu^\beta$
is a normalization constant. The parameter
$\kappa$ as well as additional parameters in the blocking function for the
blocked link $Q_\mu$
define a specific RGT and are kept fixed. They can however be
tuned for particular properties of the resulting effective
action $\beta' A'[V]$, e.g., optimizing its locality. Each RGT step decreases the resolution of the lattice,
essentially moving from left to right in Fig.~\ref{fig:continuum
  limit}, while keeping the long-distance physics intact. This is
guaranteed by choosing the normalization factor
${\cal N}_\mu^\beta$ such that $Z(\beta') = Z(\beta)$. As a
consequence, the long-distance physics at any lattice spacing $a'$ is
related to the continuum physics through the RGTs. The physics is encoded in the
effective action $\beta' A'[V]$ which, however,  involves infinitely
many couplings $\{c_\alpha'\}$.\footnote{From now on we discard the primes
  to denote RG-transformed quantities.} This is illustrated in
Fig.~\ref{fig:RGT flow}
where we show the flow of the couplings under
iterated (continuous) RGTs. Fortunately, for
asymptotically free gauge theories there is only one (marginally)
relevant coupling, namely the gauge coupling $g$, while all other
couplings $\{c_0, c_1, c_2\ldots\}$ are irrelevant. The continuum limit
$g^2 \rightarrow 0$, or equivalently $1/\beta \rightarrow 0$ as in
Fig.~\ref{fig:RGT flow}, defines the critical surface with
$\xi/a = \infty$. On that
surface the (irrelevant) couplings flow to the fixed point (FP), where they are
reproduced under the RGT,
$\{c_\alpha^\text{FP}\} \stackrel{\text{RGT}}{\longrightarrow}
\{c_\alpha^\text{FP}\}$. The universal properties in the neighbourhood
of the FP guarantee the universality of the continuum limit for
different gauge actions. Lattice actions at a very small, but finite
lattice spacing are situated just above the critical surface. Under
iterated RGTs they flow towards the renormalised trajectory (RT) which
emanates from the FP in the direction of the only relevant coupling
$g^2 \sim 1/\beta$. The couplings on the RT describe {\it quantum
  perfect actions}. These are effective actions which have no lattice artefacts at all at any
lattice spacing, because they are directly connected to the FP on
the critical surface via the RGTs. The quantum perfect actions
constitute the holy grail of Symanzik's improvement
programme: they are improved to all orders in $g^2$ and $a^2$ and a simulation at a single coarse lattice spacing reproduces exactly the
continuum long-distance physics.

Finding and constructing such an effective action is of course very
difficult and presents two practical challenges: firstly, how to 
parametrize the RT, i.e., which (finite) set of operators and
coefficients $\{c_\alpha\}$ to choose,
and secondly, how to determine the coefficients $\{c_\alpha^\text{RT}\}$ or
$\{c_\alpha^\text{FP}\}$. The first question has been answered long
ago by Hasenfratz and Niedermayer \cite{Hasenfratz:1993sp}. They
realized that for $\beta \rightarrow \infty$ (on the critical surface)
the RGT becomes a classical saddle point problem and reduces to the FP
equation
\begin{equation}
  \label{eq:FP equation}
  A^\text{FP}[V] = \min_{\{U\}} \left\{ A^\text{FP}[U] +
    T[U,V]\right\} \,.
\end{equation}
The action $A^\text{FP}$ defines an action for all values of $\beta$,
cf.~the straight line in Fig.~\ref{fig:RGT flow}. Hasenfratz and Niedermayer also realized
that the FP action is {\it classically perfect}: it has no
lattice artefacts on solutions of the classical equations of motion
and hence no tree-level lattice artefacts ${\cal O}(a^{2n})$ to all
orders in $n$. This is so because the classical FP equation
preserves all classical properties since they are connected back to the continuum
through iterations of the FP equation. The iterated FP equation essentially realises an
inception procedure for the classical properties of the gauge field configurations on coarse
lattices. In addition to the absence of tree-level lattice artefacts, the lattice
artefacts ${\cal O}(g^2 a^{2n}), n=1,2,\ldots$ induced by quantum
effects are expected to be substantially reduced because they are
suppressed by the small coupling $g^2$. This holds close to the
continuum where the FP action follows closely the RT, cf.~Fig.~\ref{fig:RGT flow}.
From the perfect classical properties it follows that the FP action has scale invariant
instanton solutions
\cite{Blatter:1995ik,DeGrand:1995ji,DeGrand:1995jk,DeGrand:1995ab}. More
importantly, it also enables a classically perfect gradient flow
as recently discussed in
Ref.~\cite{Wenger:2025sre,Holland:2025fsa}.    

In summary, the FP equation solves the first of the two practical
problems mentioned above, namely how to determine the couplings
$\{c_\alpha^\text{FP}\}$. As a solution to the second problem, namely,
how to choose a specific parametrization of the FP action in practice, in a
collaboration with Kieran Holland, Andreas Ipp and David M\"uller we have recently proposed to use
a machine-learning approach \cite{Holland:2023ews,Holland:2024muu} as described in the
next section.

\subsection{Machine learning the fixed-point action}
\begin{figure}[h]
  \includegraphics[height=0.27\textwidth]{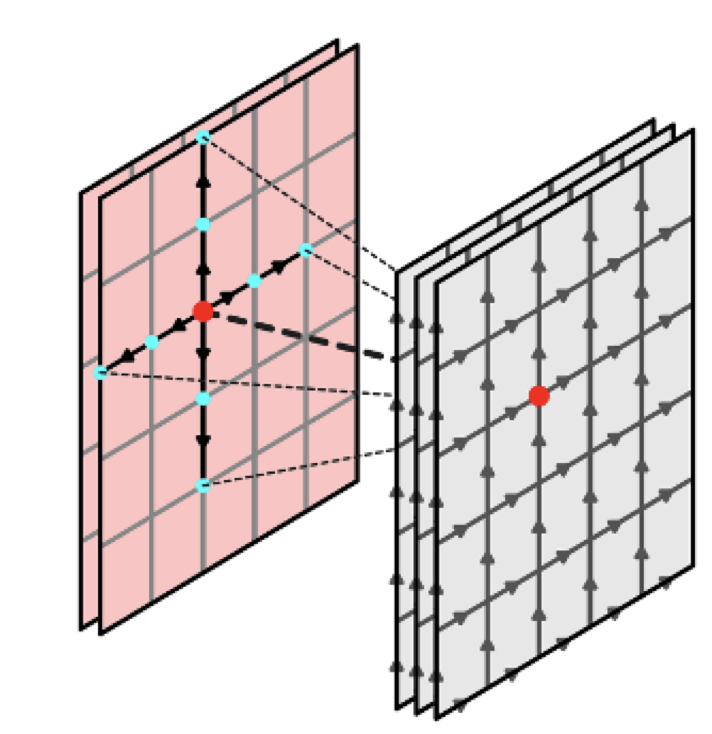}\hfill
   \includegraphics[height=0.27\textwidth]{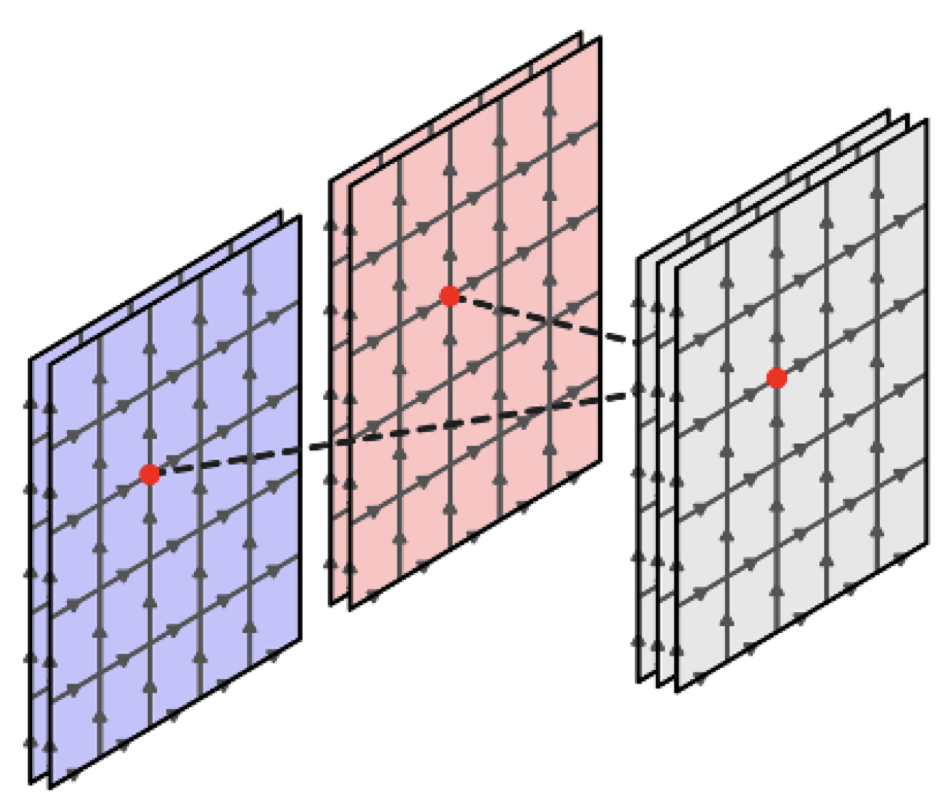}\hfill
   \includegraphics[height=0.27\textwidth]{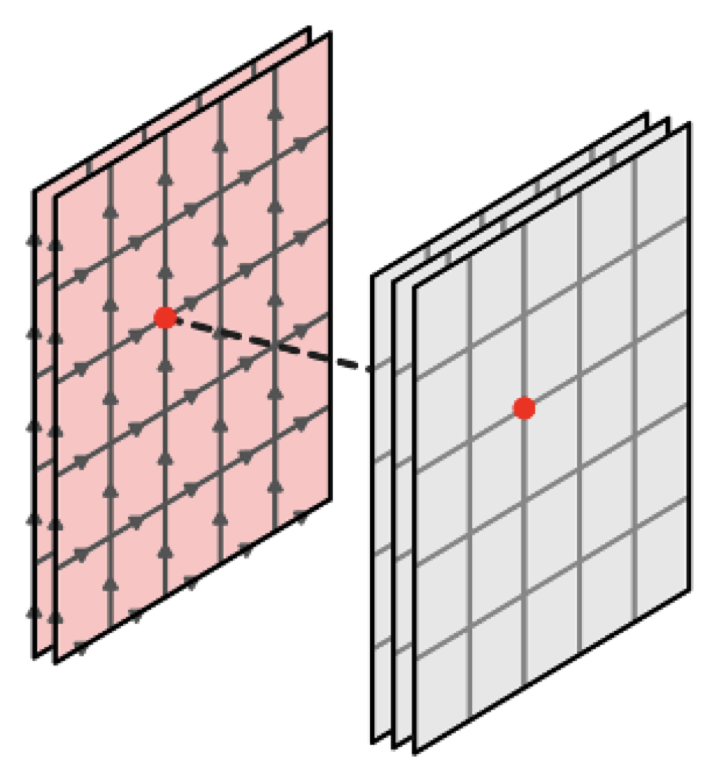} \\
   \includegraphics[width=\textwidth]{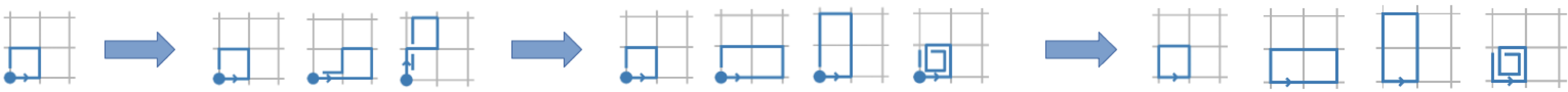}
  \caption{\label{fig:L-CNN}
    Illustrations of the L-CNN: the convolutional layer (L-Conv) in
    the left panel
    parallel transports gauge-covariant objects to a common position
    (red dot), where they are combined in
    the bilinear layer (L-Bilin) in the middle panel and eventually mapped to gauge
    invariant objects with the trace layer (L-Tr) in the right panel. The lower row
    depicts simple examples of applying the sequence of the layers above.    
  }
\end{figure}

In order to parametrize the FP action accurately and in a most
efficient way, it is crucial to choose a well-tailored
finite set of Wilson loops, together with the
corresponding coefficients $\{c_\alpha^\text{FP}\}$. In the past, a sum of powers of traces of Wilson loops has been used, starting
from a small set of simple Wilson loops, such as the plaquette and
rectangular loops
\cite{DeGrand:1995ji,DeGrand:1995jk,DeGrand:1995ab,Blatter:1996np},
and 
increasing the complexity by including smeared Wilson loops
\cite{Niedermayer:2000yx}. More recently, it has been proposed in
Ref.~\cite{Holland:2023ews,Holland:2024muu} to use a machine-learning approach. It 
employs a lattice gauge equivariant convolutional neural network
(L-CNN) which is capable of generating arbitrarily shaped Wilson loops
in a systematic way \cite{Favoni:2020reg}. The coefficients of these loops can then be
machine learned to reproduce arbitrary gauge equivariant or gauge invariant functions of the
gauge fields. The main ingredients of the network are illustrated in
Fig.~\ref{fig:L-CNN}.
The lattice-convolutional layers (L-Conv) take
gauge covariant objects
$W$ and parallel transport them from positions $x + k \hat \mu$ to $x$
using the gauge links $U$ producing new, gauge covariant objects
$W'_{x,i} = \sum_{j,\mu,k} w_{i,j,\mu,k} U_{x,k\cdot \hat \mu} W_{x+k\cdot \hat
  \mu,j} U^\dagger_{x,k\cdot \hat \mu}$ where $U_{x,k\cdot \hat
  \mu}=U_{x,\mu} \cdot \ldots \cdot U_{x+{k-1}\cdot \hat \mu}$. The lattice-bilinear layers (L-Bilin) take two gauge covariant objects
$W$ and $W'$ and produce bilinear combinations $W''_{x,i} = \sum_{j,j'}
\alpha_{i,j,j'} W_{x,j} W'_{x, j'}$. The trace
layer (L-Tr) eventually produces gauge invariant objects $w_{x,i} = \text{Tr } W_{x,j}
\in \mathbb{C}$ and the network can be further enhanced by
supplementing additional activation layers. The ranges of the indices
$i,j,k$ are part of the architecture choice, while the coefficients $\{w_{i,j,\mu,k}\}$ and
$\{\alpha_{i,j,j'}\}$ are the trainable parameters independent of $x$
making the network translationally
invariant.

The data set for the supervised machine learning of the trainable parameters is
produced as follows.
A large range of gauge field ensembles are generated with the Wilson
gauge action (or any other action) covering field
fluctuations from very smooth to very coarse. For each configuration $V$
one then solves the FP Eq.~(\ref{eq:FP equation}) by finding the
minimizing configuration $U_\text{min}$ on the RHS
using a sufficiently good approximation of $A^\text{FP}[U]$, e.g., as
provided in \cite{Niedermayer:2000yx}. This yields the exact FP action
value $A^\text{FP}[V]$ and in addition also the exact derivatives w.r.t.~to
each gauge link,
\begin{equation}
  \label{eq:FP derivatives}
\frac{\delta A^\text{FP}[V]}{\delta V^a_{x,\mu}} = \frac{\delta
  T[U_\text{min},V]} {\delta V^a_{x,\mu}} = -\kappa \, \text{Re Tr}(i
t^a \,V_{x,\mu} Q^\dagger_{x,\mu}[U_\text{min}]) \, ,
\end{equation}
where $x$ runs over all sites of the coarse lattice, $\mu=1,\ldots, 4$
over all spacetime directions,
and $a=1,\ldots,8$ over the colour indices. In total, this yields
$4\times 8\times (\text{volume}) + 1$ learning data per
configuration. The direct parametrization of the derivatives is of
course most useful for simulating the action with the HMC algorithm,
or for constructing the gradient flow.  

\begin{figure}[t]
  \vspace*{-0.5cm}
     \includegraphics[width=\textwidth]{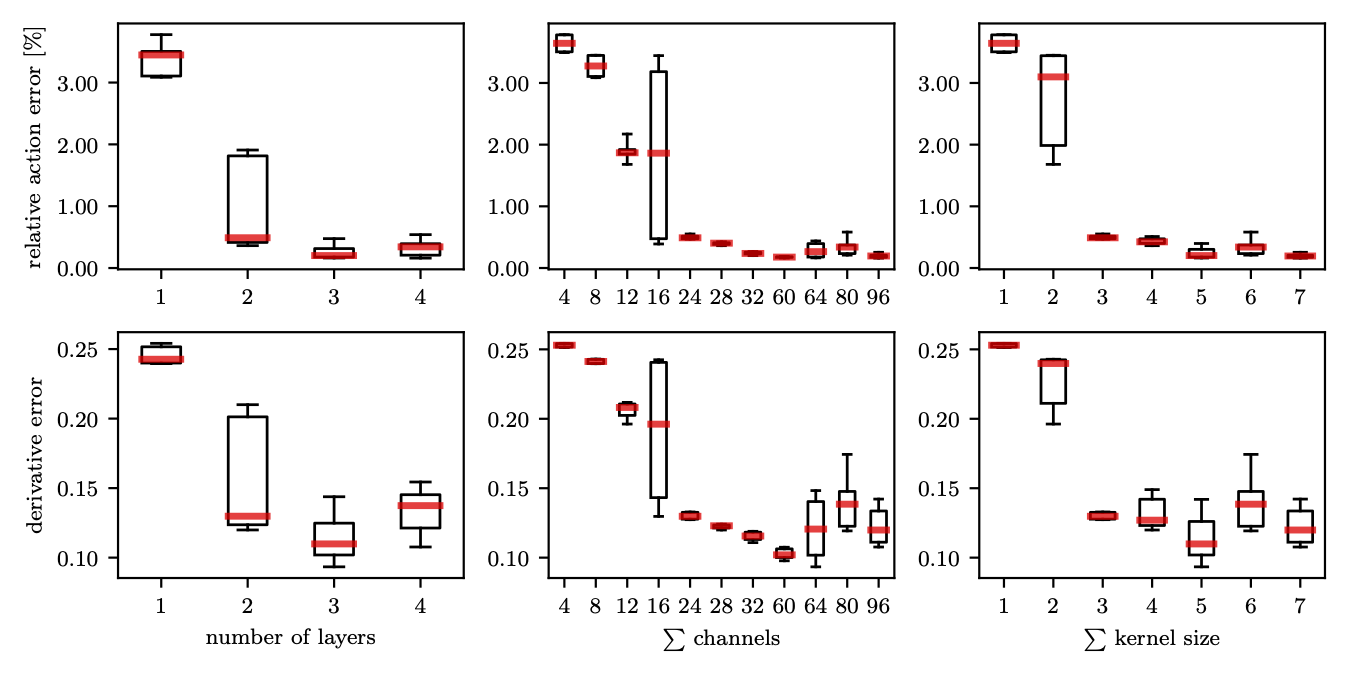}
  \caption{\label{fig:architecture search}
    Illustrations of the architecture search. Shown are the relative
    action error ({\it top row}) and derivative error ({\it bottom row}) of the
    L-CNN w.r.t.~to the exact FP values as a function of the
    architecture parameters, i.e., the number of layers (left plots),
    the total number of channels ({\it middle plots}), and the kernel size
    ({\it right plots}).
    \vspace{-0.3cm}
  }
\end{figure}

An important step in any machine-learning approach is to find an appropriate network
architecture. The outcome of such a search for an optimal achitecture is
shown in Fig.~\ref{fig:architecture search} where the distributions of
the relative action errors and the derivative errors of the L-CNN
output w.r.t.~to the exact FP values in Eqs.~(\ref{eq:FP equation})
and (\ref{eq:FP derivatives}) are represented in terms of box
plots.  
From the dependence on the number of combined convolutional and
bilinear layer pairs (left row), the total number of output channels
(middle row) and kernel sizes (right row) one can identify good
architectures. The one chosen in Ref.~\cite{Holland:2024muu} consists
of three combined layers with 12, 24, 24 output channels and kernel
sizes $k=2,2,1$, respectively, yielding a total of about 443k
parameters.  Since the number of layers is finite, the L-CNN produces
an ultralocal action and therefore a truncated
approximation of the FP action. However, seeing that 
the effective couplings between two
gauge links in the parametrized
action decay exponentially fast, at least as $\exp(-3.13 r)$ with
their separation $r$, cf.~\cite{Holland:2024muu}, the truncation error
is expected to be small. In any case, what matters for the successful
description of the FP action is of course the
total parametrization error. Eventually, the quality of the parametrization can
only be checked in actual simulations.

\subsection{The fixed-point action in action}
\begin{figure}[t]
  \vspace*{-0.5cm}
     \includegraphics[width=0.5\textwidth]{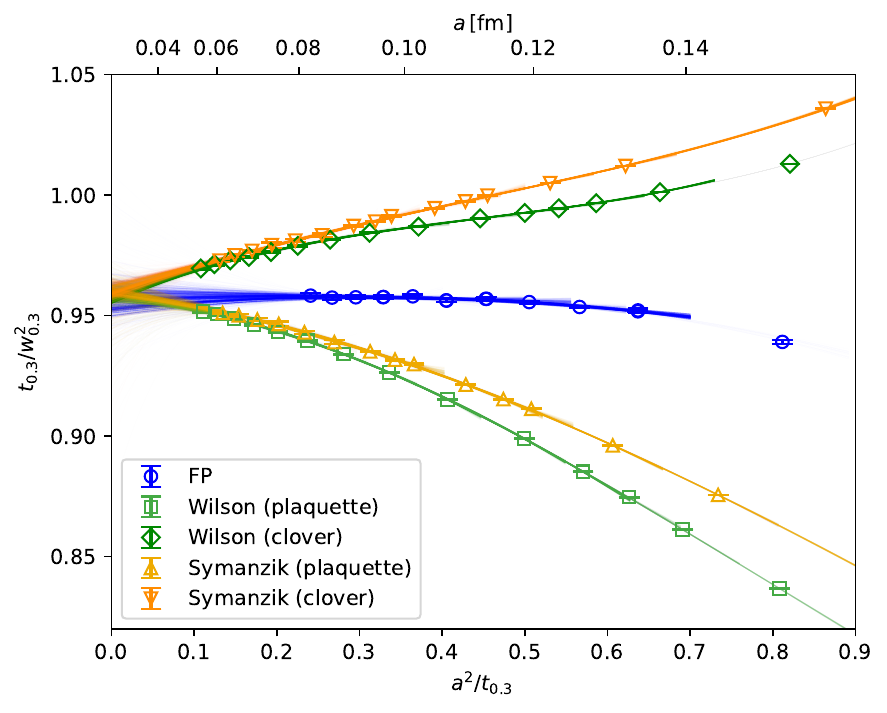}
     \includegraphics[width=0.5\textwidth]{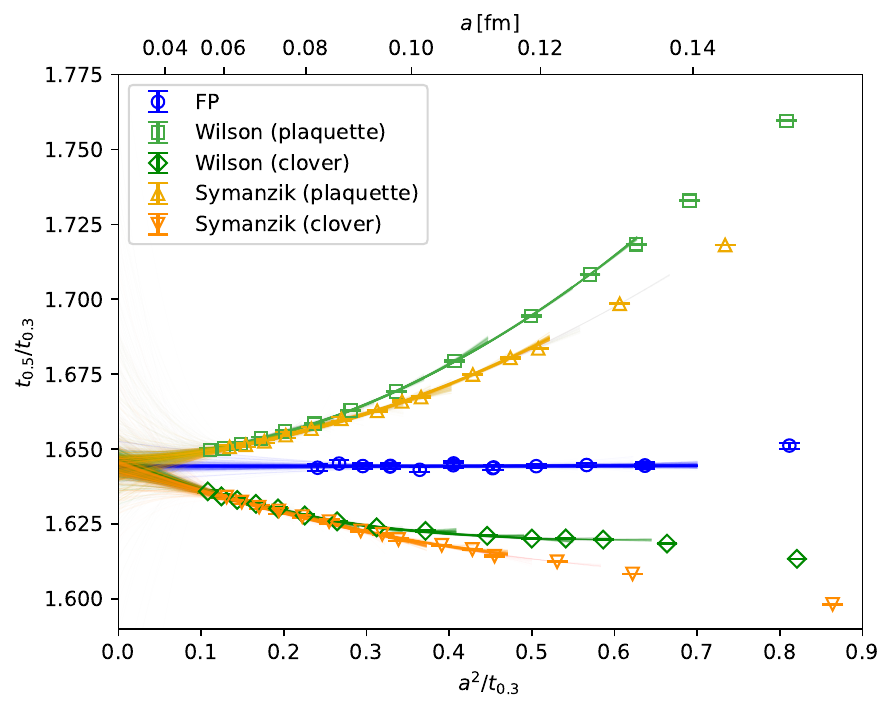}
  \caption{\label{fig:GF scaling}
    Continuum-limit extrapolations for the ratios
$t_{0.3}/w^2_{
0.3}$ and $t_{0.5}/t_{0.3}$. Results from Wilson and Symanzik
MC simulations are shown using plaquette and clover discretizations of
the action density.
\vspace{-0.3cm}
  }
\end{figure}

While the FP action has a very complicated structure in terms of a
plethora of extended Wilson loops, the derivatives with respect to the gauge
links are directly obtained as an output from the L-CNN. As
a consequence, HMC simulations and the calculations of gradient-flow observables
are straightforward. In fact, the gradient flow provides an ideal test case for
the FP approach and
the quality of the machine-learned parametrization of the FP action,
because gradient-flow observables can be measured very precisely and entail
essentially no systematic errors. They are therefore ideal candidates
to quantify lattice artefacts and test the scaling towards the
continuum limit. In particular, one can consider physical reference
scales $t_x$ and $w_x$  \cite{Luscher:2010iy,BMW:2012hcm} defined by
\begin{equation}
  t^2 \langle E \rangle |_{t={t_{x}}}=x, \qquad t \frac{d}{dt} \left( t^2 \langle E \rangle \right) \big|_{t=w_{x}^2} = x
\end{equation}
yielding dimensionless ratios $t_x/w_x^2$ or $t_x/t_y$ as scaling
quantities with well defined continuum limits. Because the FP action is classically
perfect, no lattice artefacts are introduced through the gradient flow or the
measurement of the action density $E(t)$, and the only deviations from
perfect scaling are either due to quantum lattice artefacts of order
$O(g^2 a^{2n})$, or due to the imperfect parametrization of the FP action by the machine-learned L-CNN. The results for the dimensionless ratios
$t_{0.5}/t_{0.3}$ (left plot) and $t_{0.3}/w_{0.3}$ (right plot) as a
function of the lattice spacing $a^2/t_{0.3}$ are shown in
Fig.~\ref{fig:GF scaling} using MC simulations of the FP,
Wilson and the tree-level Symanzik-improved gauge actions. For the
latter two, the action density is evaluated with 
either the plaquette or clover operator. The leading lattice
artefacts are expected to be $O(a^2)$ for the Wilson data,
$O(a^4)$ for the Symanzik data, and $O(g^2 a^2)$ for the FP
data. The lines represent bootstrap samples of various possible fits
describing the continuum extrapolations. We note that the leading $O(a^4)$ behaviour of the Symanzik data
is masked by the $O(a^2)$ discretization effects from the flow action
(Wilson) and the action density (plaquette and clover). This can be
remedied by employing the Zeuthen flow introduced in
Ref.~\cite{Ramos:2015baa} yielding $O(a^2)$ improved gradient-flow
observables. In contrast, the improvement with the machine-learned
FP action is to all orders in the lattice spacing at tree
level.
Indeed, the observed lattice artefacts for the FP action are
less than 1\% up to lattice spacings of 0.14 fm, allowing continuum
physics to be extracted for SU(3) gauge theory in four spacetime
dimensions from coarse lattices.
\begin{figure}[b]
     \includegraphics[width=0.5\textwidth]{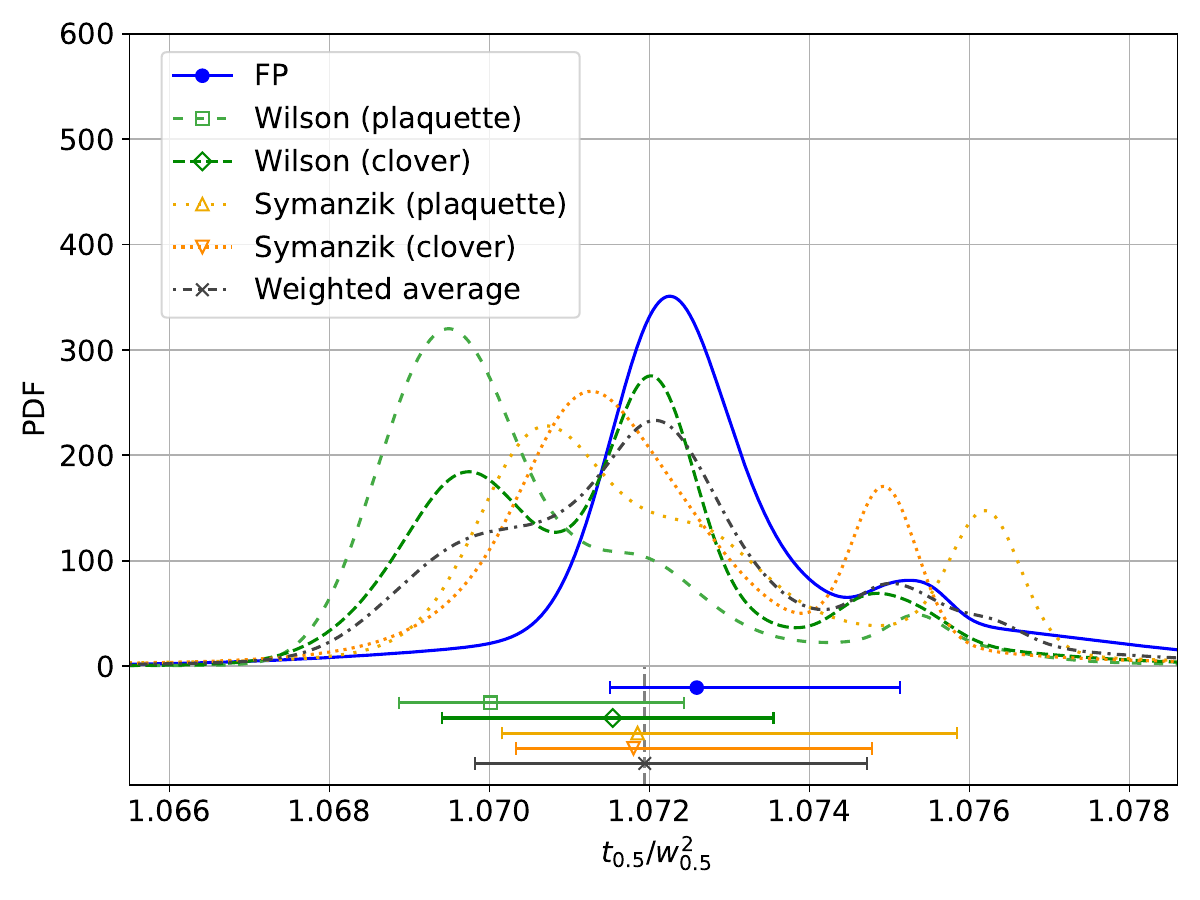}
     \includegraphics[width=0.5\textwidth]{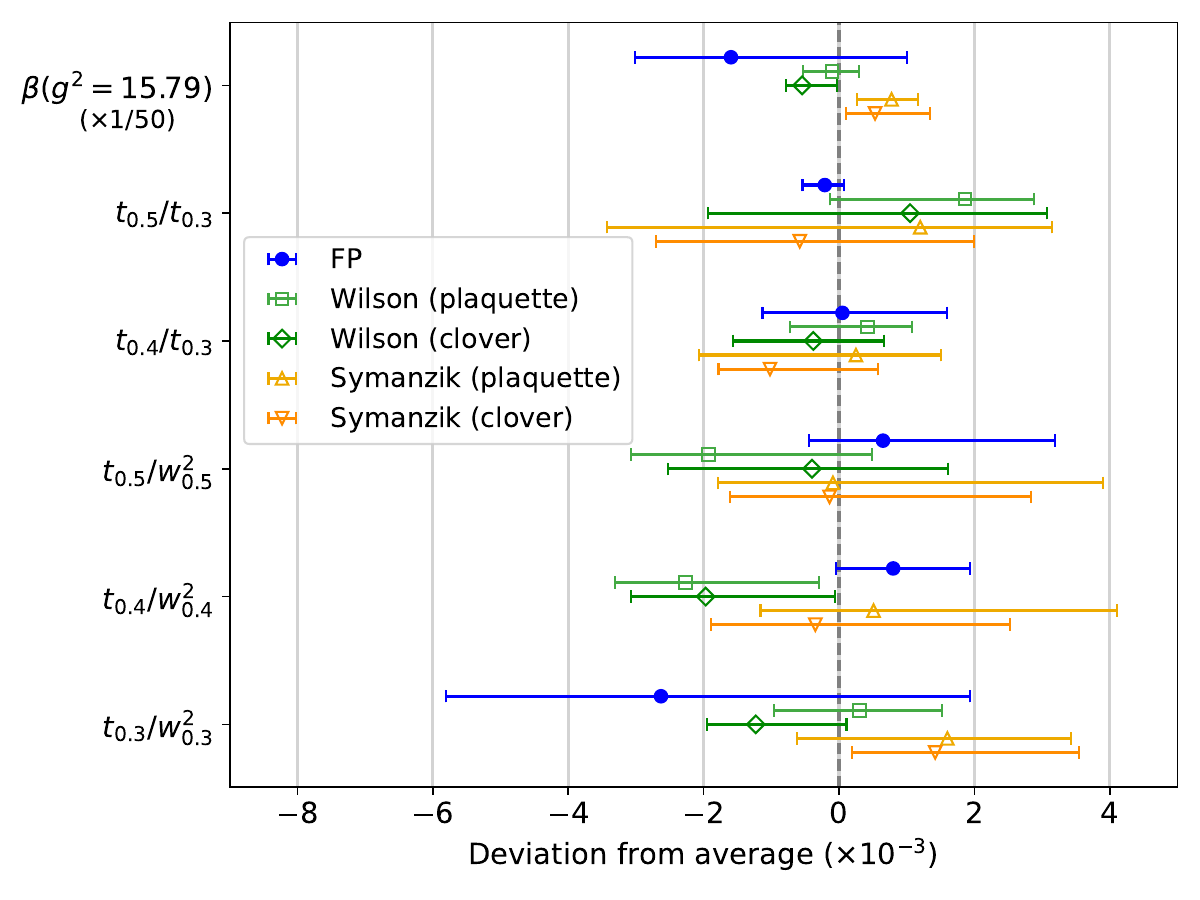}
  \caption{\label{fig:summary of continuum limits}
     {\it Left plot:} Comparison of various PDFs obtained from AIC-weighted
     continuum limits of the ratio $t_{0.5}/w_{0.5}^2$ using a variety of fit functions and
     ranges. {\it Right plot:} Comparison of continuum predictions for four-dimensional SU(3) gauge theory from MC simulations using
either the FP, Wilson or tree-level Symanzik improved lattice
action. The $\beta$-function results
are rescaled by a factor of 50 for visibility.}
\end{figure}
 This is further exemplified in
Fig.~\ref{fig:summary of continuum limits} where the plot on the left
shows the  AIC-weighted PDFs of the continuum limits for the ratio
$t_{0.5}/w^2_{0.5}$ using a variety
of fit functions and ranges for the FP, Wilson and Symanzik action. The plot shows how the continuum values of the gradient-flow
observables are dominated by the systematic effects from the continuum
extrapolations. It underlines the importance of controlling lattice
artefacts well. The right plot shows a comparison of continuum
predictions for a variety of  gradient-flow observables, including the $\beta$-function at the
gradient-flow coupling $g^2=15.78$. There is clear consistency between FP, Wilson
and Symanzik results demonstrating universality and the successful implementation of
the machine-learned L-CNN for the FP approach. 

The classically perfect improvement is not limited to gradient-flow
observables, but in fact also applies to spectral observables. This is
illustrated in the left plot of Fig.~\ref{fig:static potential and
  deconfinement transition} where results for the static
\begin{figure}[t]
     \includegraphics[width=0.48\textwidth,valign=t]{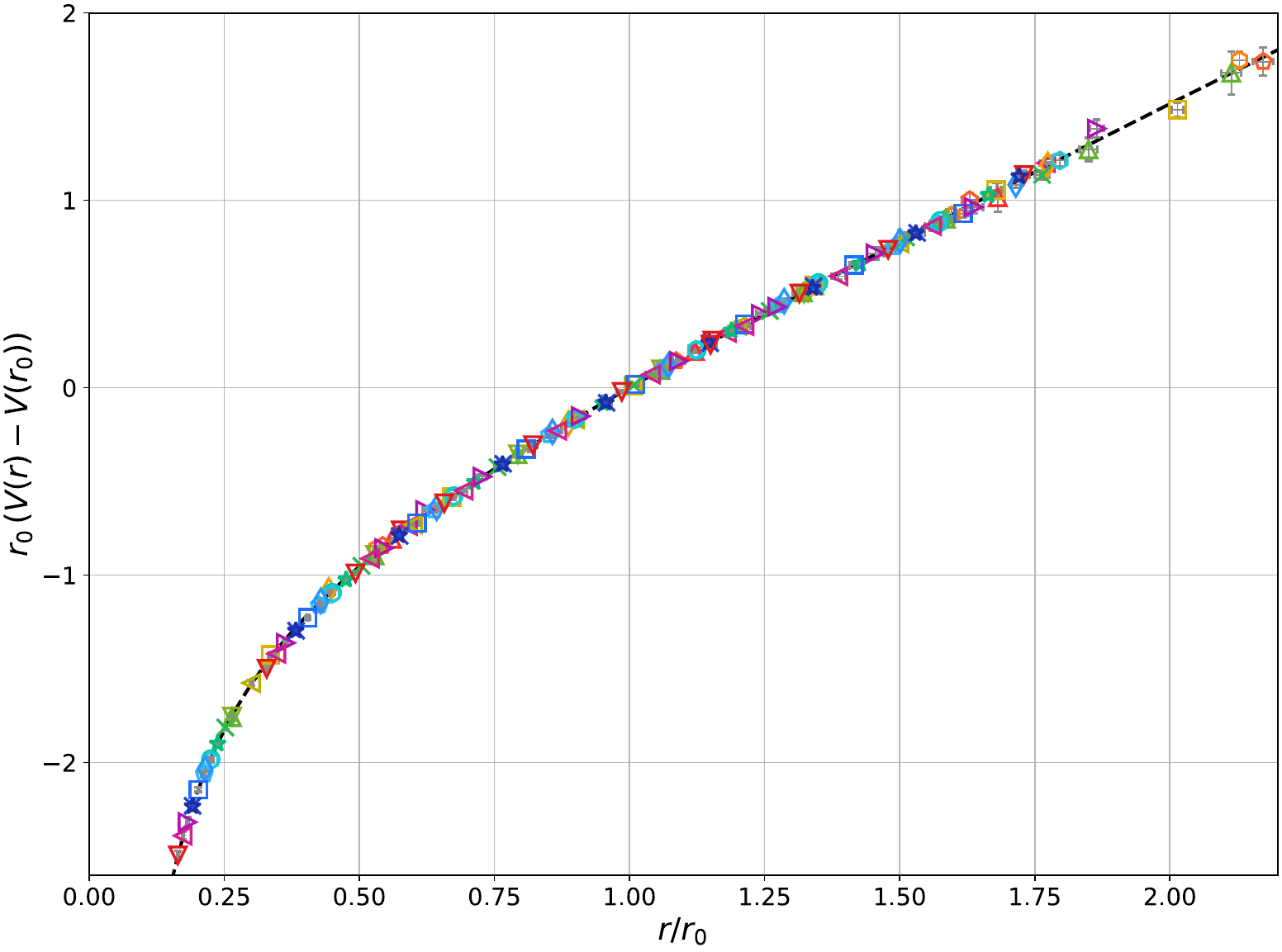}
     \includegraphics[width=0.5\textwidth,valign=t]{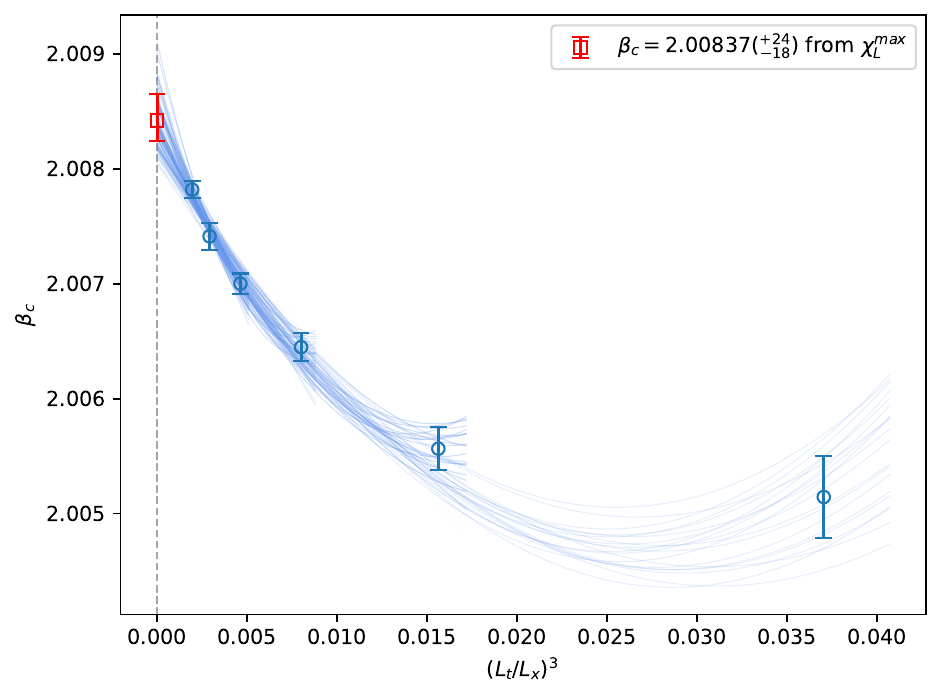}
  \caption{\label{fig:static potential and deconfinement transition}
     {\it Left plot:} Static quark-antiquark potential obtained on lattices
     with lattice spacings ranging between 0.08 fm $\leq a \leq 0.3$
     fm. {\it Right plot:} Thermodynamic limit of the critical coupling
     $\beta_c$ obtained from the position of the Polyakov loop
     susceptibility at a lattice spacing of $a\simeq 0.3$ fm
     corresponding to a temporal lattice extent of $L_t=2$.}
\end{figure}
quark-antiquark potential are shown including simulations at lattice
spacings as coarse as $a\simeq 0.3$ fm. There are essentially no
lattice artefacts visible even at this coarse resolution. In contrast,
the right
plot is not about lattice artefacts, but rather
about finite-volume effects. It shows the extrapolation of the
critical coupling $\beta_c$ to the thermodynamic limit. The critical
coupling is obtained from the position of the peak of the Polyakov
loop susceptibility at a lattice spacing of
$a\simeq 0.3$ fm corresponding to a temporal lattice extent of
$L_t=2$. Simulating at such a coarse lattice
spacing has the advantage that it is particularly easy to reach large
aspect ratios. As a consequence, also in this situation the FP approach is most
suited and the machine-learned FP action indeed appears
to perform well.

\section{Summary and conclusions}
Since the application of machine-learning techniques in lattice gauge theories is a rather new
field, many developments are still in an exploratory state. Hence,
there are many opportunities for making progress. It is
interesting to see the many creative combinations of often complementary
approaches and tools that are employed. There is a large variety of different ideas, many
of which will probably not be successful in the end, but
nevertheless need to be tried out. 

The efforts of applying machine learning for lattice gauge theories
are often motivated by the urge to overcome problems related to
critical slowing down towards the continuum, in particular topological
freezing.  One lesson to be learned from this brief review is that it
is in general not sufficient to just adapt a given machine-learned
generative model to a gauge theory, but instead it requires the
employment of additional physics-based concepts in order to enable
efficient sampling of gauge field configurations.
Examples for such physics-driven machine-learning approaches are the
physics-conditioned diffusion models \cite{Zhu:2025pmw}, the
stochastic-normalizing-flow models enhanced by non-equilibrium
dynamics \cite{Bulgarelli:2024brv,Bonanno:2025pdp}, and the
machine-learned FP actions based on renormalization group
transformations \cite{Holland:2024muu,Holland:2025fsa}, as discussed
in this review.
Another interesting proposal in that direction has been put forward
very recently in
Ref.~\cite{Ihssen:2025ybn} where physics-informed renormalization
group flows are used to transform the generative process into that of solving
differential equations for the kernels of the RG transformation.
The necessity of including physics information into
the machine learning may be a hint why approaches solely based on
normalizing flows without any physics input have not yet been able to
scale to realistic lattice volumes in four spacetime dimensions.

In fact, it is probably safe to state that the step from low-dimensional applications
with simple degrees of freedom to four-dimensional large-volume
applications is most challenging. So far, not many applications have
sucessfully mastered this step, with the notable exceptions being
\cite{Bulgarelli:2024brv,Bonanno:2025pdp} and
\cite{Holland:2024muu,Holland:2025fsa}. However, given the variety of ideas
available, there is no doubt that there will soon be many more
approaches 
establishing machine learning as a transformative tool for overcoming
computational challenges in lattice gauge theories.

\section*{Acknowledgments}
I would like to thank K.~Holland, A.~Ipp, and D.I.~M\"uller for a most enjoyable collaboration.
I would also like to thank G.~Aarts, L.~Backfried, C.~Bonanno,
E.~Cellini, G.~Kanwar, J.~Mayer-Steudte, A.~Nada, F.~Romero-L\'opez,
S.~Romiti, and L.~Verzichelli for useful discussions, and S.~Thommen for providing the data in the left plot in Fig.~\ref{fig:static potential and deconfinement transition}.

This work has been supported by the Swiss
National Science Foundation (SNSF) through grant No.~200020\_208222 and
No.~200021-232288, the Platform for Advanced Scientific Computing
(PASC) under the project Alpenglue, and the Albert Einstein Center for Fundamental
Physics at the University of Bern. The computational results presented
in the context of the FP action have been achieved in part using
the Vienna Scientific Cluster (VSC) and LEONARDO at CINECA, Italy, via
an AURELEO (Austrian Users at LEONARDO supercomputer) project, and UBELIX, the
HPC cluster at the University of Bern.

\bibliographystyle{JHEP}
\bibliography{references} 

@article{Abbott:2026ylv,
    author = "Abbott, Ryan and Boyda, Denis and Fu, Yang and Hackett, Daniel C. and Kanwar, Gurtej and Romero-L{\'o}pez, Fernando and Shanahan, Phiala E. and Urban, Julian M.",
    title = "{Variance reduction in lattice QCD observables via normalizing flows}",
    eprint = "2603.02984",
    archivePrefix = "arXiv",
    primaryClass = "hep-lat",
    reportNumber = "FERMILAB-PUB-26-0130-T, MIT-CTP/6010",
    month = "3",
    year = "2026"
}

@article{Aarts:2026zzr,
    author = {Aarts, Gert and Habibi, Diaa E. and Ipp, Andreas and M{\"u}ller, David I. and Ranner, Thomas R. and Wang, Lingxiao and Wang, Wei and Zhu, Qianteng},
    title = "{Generalizable Equivariant Diffusion Models for Non-Abelian Lattice Gauge Theory}",
    eprint = "2601.19552",
    archivePrefix = "arXiv",
    primaryClass = "hep-lat",
    reportNumber = "RIKEN-iTHEMS-Report-26",
    month = "1",
    year = "2026"
}

@article{Bonanno:2020hht,
    author = "Bonanno, Claudio and Bonati, Claudio and D'Elia, Massimo",
    title = "{Large-$N$ $SU(N)$ Yang-Mills theories with milder topological freezing}",
    eprint = "2012.14000",
    archivePrefix = "arXiv",
    primaryClass = "hep-lat",
    doi = "10.1007/JHEP03(2021)111",
    journal = "JHEP",
    volume = "03",
    pages = "111",
    year = "2021"
}

@inproceedings{Bonanno:2026tle,
    author = "Bonanno, Claudio and Bulgarelli, Andrea and Cellini, Elia and Nada, Alessandro and Panfalone, Dario and Vadacchino, Davide and Verzichelli, Lorenzo",
    title = "{A scalable flow-based approach to mitigate topological freezing}",
    booktitle = "{42th International Symposium on Lattice Field Theory}",
    eprint = "2601.20708",
    archivePrefix = "arXiv",
    primaryClass = "hep-lat",
    month = "1",
    year = "2026"
}

@article{Bellscheidt:2026rjh,
    author = "Bellscheidt, Verena and Brambilla, Nora and Kronfeld, Andreas S. and Mayer-Steudte, Julian",
    title = "{Wilson loops with neural networks}",
    eprint = "2602.02436",
    archivePrefix = "arXiv",
    primaryClass = "hep-lat",
    reportNumber = "TUM-EFT 202/25, FERMILAB-PUB-26-0041-T, MIT-CTP/5995",
    month = "2",
    year = "2026"
}

@article{Ihssen:2025ybn,
    author = "Ihssen, Friederike and Kapust, Renzo and Pawlowski, Jan M.",
    title = "{Generative sampling with physics-informed kernels}",
    eprint = "2510.26678",
    archivePrefix = "arXiv",
    primaryClass = "hep-lat",
    month = "10",
    year = "2025"
}

@article{Lawrence:2025rnk,
    author = "Lawrence, Scott",
    title = "{Machine-learning approaches to accelerating lattice simulations}",
    eprint = "2502.02670",
    archivePrefix = "arXiv",
    primaryClass = "hep-lat",
    reportNumber = "LA-UR-25-20975",
    doi = "10.22323/1.466.0010",
    journal = "PoS",
    volume = "LATTICE2024",
    pages = "010",
    year = "2025"
}

@inproceedings{Detmold:2024mts,
    author = "Detmold, William and Kanwar, Gurtej and Lin, Yin and Shanahan, Phiala E. and Wagman, Michael L.",
    title = "{Exploring gauge-fixing conditions with gradient-based optimization}",
    booktitle = "{41st International Symposium on Lattice Field Theory}",
    eprint = "2410.03602",
    archivePrefix = "arXiv",
    primaryClass = "hep-lat",
    reportNumber = "MIT-CTP/5786, FERMILAB-CONF-24-0732-T",
    month = "10",
    year = "2024"
}

@article{Romiti:2025cfs,
    author = "Romiti, Simone",
    title = "{SU(N) lattice gauge theories with physics-informed neural networks}",
    eprint = "2510.26904",
    archivePrefix = "arXiv",
    primaryClass = "hep-lat",
    doi = "10.1103/mb67-9rkf",
    journal = "Phys. Rev. D",
    volume = "113",
    number = "5",
    pages = "054511",
    year = "2026"
}

@inproceedings{Kanwar:2024ujc,
    author = "Kanwar, Gurtej",
    title = "{Flow-based sampling for lattice field theories}",
    booktitle = "{40th International Symposium on Lattice Field Theory}",
    eprint = "2401.01297",
    archivePrefix = "arXiv",
    primaryClass = "hep-lat",
    month = "1",
    year = "2024"
}

@article{Abbott:2024mix,
    author = "Abbott, Ryan and Boyda, Denis and Hackett, Daniel C. and Kanwar, Gurtej and Romero-L{\'o}pez, Fernando and Shanahan, Phiala E. and Urban, Julian M. and Albergo, Michael S.",
    title = "{Practical applications of machine-learned flows on gauge fields}",
    eprint = "2404.11674",
    archivePrefix = "arXiv",
    primaryClass = "hep-lat",
    reportNumber = "FERMILAB-CONF-24-0007-T, MIT-CTP/5669",
    doi = "10.22323/1.453.0011",
    journal = "PoS",
    volume = "LATTICE2023",
    pages = "011",
    year = "2024"
}

@article{Kanwar:2020xzo,
    author = "Kanwar, Gurtej and Albergo, Michael S. and Boyda, Denis and Cranmer, Kyle and Hackett, Daniel C. and Racani{\`e}re, S{\'e}bastien and Rezende, Danilo Jimenez and Shanahan, Phiala E.",
    title = "{Equivariant flow-based sampling for lattice gauge theory}",
    eprint = "2003.06413",
    archivePrefix = "arXiv",
    primaryClass = "hep-lat",
    reportNumber = "MIT-CTP/5181",
    doi = "10.1103/PhysRevLett.125.121601",
    journal = "Phys. Rev. Lett.",
    volume = "125",
    number = "12",
    pages = "121601",
    year = "2020"
}

@article{Boyda:2020hsi,
    author = "Boyda, Denis and Kanwar, Gurtej and Racani{\`e}re, S{\'e}bastien and Rezende, Danilo Jimenez and Albergo, Michael S. and Cranmer, Kyle and Hackett, Daniel C. and Shanahan, Phiala E.",
    title = "{Sampling using $SU(N)$ gauge equivariant flows}",
    eprint = "2008.05456",
    archivePrefix = "arXiv",
    primaryClass = "hep-lat",
    reportNumber = "MIT-CTP/5228",
    doi = "10.1103/PhysRevD.103.074504",
    journal = "Phys. Rev. D",
    volume = "103",
    number = "7",
    pages = "074504",
    year = "2021"
}

@article{Abbott:2023thq,
    author = "Abbott, Ryan and others",
    title = "{Normalizing flows for lattice gauge theory in arbitrary space-time dimension}",
    eprint = "2305.02402",
    archivePrefix = "arXiv",
    primaryClass = "hep-lat",
    month = "5",
    year = "2023"
}

@article{Abbott:2024kfc,
    author = "Abbott, Ryan and Botev, Aleksandar and Boyda, Denis and Hackett, Daniel C. and Kanwar, Gurtej and Racani{\`e}re, S{\'e}bastien and Rezende, Danilo J. and Romero-L{\'o}pez, Fernando and Shanahan, Phiala E. and Urban, Julian M.",
    title = "{Applications of flow models to the generation of correlated lattice QCD ensembles}",
    eprint = "2401.10874",
    archivePrefix = "arXiv",
    primaryClass = "hep-lat",
    reportNumber = "MIT-CTP/5658, FERMILAB-PUB-24-0014-T",
    doi = "10.1103/PhysRevD.109.094514",
    journal = "Phys. Rev. D",
    volume = "109",
    number = "9",
    pages = "094514",
    year = "2024"
}

@article{Abbott:2025kvi,
    author = "Abbott, Ryan and Boyda, Denis and Kanwar, Gurtej and Romero-L{\'o}pez, Fernando and Hackett, Daniel C. and Shanahan, Phiala E. and Urban, Julian M.",
    title = "{Progress in Normalizing Flows for 4d Gauge Theories}",
    eprint = "2502.00263",
    archivePrefix = "arXiv",
    primaryClass = "hep-lat",
    reportNumber = "MIT-CTP/5839, FERMILAB-CONF-25-0049-T",
    doi = "10.22323/1.466.0066",
    journal = "PoS",
    volume = "LATTICE2024",
    pages = "066",
    year = "2025"
}

@misc{lai2025principlesdiffusionmodels,
      title={The Principles of Diffusion Models}, 
      author={Chieh-Hsin Lai and Yang Song and Dongjun Kim and Yuki Mitsufuji and Stefano Ermon},
      year={2025},
      eprint={2510.21890},
      archivePrefix={arXiv},
      primaryClass={cs.LG},
      url={https://arxiv.org/abs/2510.21890}, 
}

@article{Wang:2023exq,
    author = "Wang, Lingxiao and Aarts, Gert and Zhou, Kai",
    title = "{Diffusion models as stochastic quantization in lattice field theory}",
    eprint = "2309.17082",
    archivePrefix = "arXiv",
    primaryClass = "hep-lat",
    reportNumber = "RIKEN-iTHEMS-Report-24",
    doi = "10.1007/JHEP05(2024)060",
    journal = "JHEP",
    volume = "05",
    pages = "060",
    year = "2024"
}

@article{Zhu:2025pmw,
    author = "Zhu, Qianteng and Aarts, Gert and Wang, Wei and Zhou, Kai and Wang, Lingxiao",
    title = "{Physics-conditioned diffusion models for lattice gauge theory}",
    eprint = "2502.05504",
    archivePrefix = "arXiv",
    primaryClass = "hep-lat",
    reportNumber = "RIKEN-iTHEMS-Report-25",
    doi = "10.1007/JHEP03(2026)111",
    journal = "JHEP",
    volume = "03",
    pages = "111",
    year = "2026"
}

@article{Aarts:2025lpi,
    author = "Aarts, Gert and Habibi, Diaa E. and Wang, Lingxiao and Zhou, Kai",
    title = "{Combining complex Langevin dynamics with score-based and energy-based diffusion models}",
    eprint = "2510.01328",
    archivePrefix = "arXiv",
    primaryClass = "hep-lat",
    reportNumber = "RIKEN-iTHEMS-Report-25",
    doi = "10.1007/JHEP12(2025)160",
    journal = "JHEP",
    volume = "12",
    pages = "160",
    year = "2025"
}

@article{Alharazin:2026lcb,
    author = "Alharazin, H. and Panteleeva, J. Yu. and Sun, B. -D.",
    title = "{Diffusion Models for SU(2) Lattice Gauge Theory in Two Dimensions}",
    eprint = "2602.09045",
    archivePrefix = "arXiv",
    primaryClass = "hep-lat",
    month = "2",
    year = "2026"
}

@article{Fukushima:2024oij,
    author = "Fukushima, Kenji and Kamata, Syo and Hirono, Yuji",
    title = "{Stochastic Quantization and Diffusion Models}",
    eprint = "2411.11297",
    archivePrefix = "arXiv",
    primaryClass = "hep-lat",
    doi = "10.7566/JPSJ.94.031010",
    journal = "J. Phys. Soc. Jap.",
    volume = "94",
    number = "3",
    pages = "031010",
    year = "2025"
}

@article{Vega:2025hgz,
    author = "Vega, Octavio and Komijani, Javad and El-Khadra, Aida and Marinkovic, Marina",
    title = "{Group-Equivariant Diffusion Models for Lattice Field Theory}",
    eprint = "2510.26081",
    archivePrefix = "arXiv",
    primaryClass = "hep-lat",
    month = "10",
    year = "2025"
}

@inproceedings{Kanwar:2025wuc,
    author = "Kanwar, Gurtej and Vega, Octavio",
    title = "{Spectral Diffusion for Sampling on ${\rm SU}(N)$}",
    booktitle = "{42th International Symposium on Lattice Field Theory}",
    eprint = "2512.19877",
    archivePrefix = "arXiv",
    primaryClass = "hep-lat",
    month = "12",
    year = "2025"
}

@article{Bonanno:2024udh,
    author = "Bonanno, Claudio and Nada, Alessandro and Vadacchino, Davide",
    title = "{Mitigating topological freezing using out-of-equilibrium simulations}",
    eprint = "2402.06561",
    archivePrefix = "arXiv",
    primaryClass = "hep-lat",
    doi = "10.1007/JHEP04(2024)126",
    journal = "JHEP",
    volume = "04",
    pages = "126",
    year = "2024"
}

@article{Bulgarelli:2024cqc,
    author = "Bulgarelli, Andrea and Cellini, Elia and Nada, Alessandro",
    title = "{Sampling SU(3) pure gauge theory with Stochastic Normalizing Flows}",
    eprint = "2409.18861",
    archivePrefix = "arXiv",
    primaryClass = "hep-lat",
    doi = "10.22323/1.466.0040",
    journal = "PoS",
    volume = "LATTICE2024",
    pages = "040",
    year = "2025"
}

@article{Bulgarelli:2024yrz,
    author = {Bulgarelli, Andrea and Cellini, Elia and Jansen, Karl and K{\"u}hn, Stefan and Nada, Alessandro and Nakajima, Shinichi and Nicoli, Kim A. and Panero, Marco},
    title = "{Flow-Based Sampling for Entanglement Entropy and the Machine Learning of Defects}",
    eprint = "2410.14466",
    archivePrefix = "arXiv",
    primaryClass = "quant-ph",
    doi = "10.1103/PhysRevLett.134.151601",
    journal = "Phys. Rev. Lett.",
    volume = "134",
    number = "15",
    pages = "151601",
    year = "2025"
}

@article{Vadacchino:2024lob,
    author = "Vadacchino, Davide and Nada, Alessandro and Bonanno, Claudio",
    title = "{Topological susceptibility of SU(3) pure-gauge theory from out-of-equilibrium simulations}",
    eprint = "2411.00620",
    archivePrefix = "arXiv",
    primaryClass = "hep-lat",
    doi = "10.22323/1.466.0415",
    journal = "PoS",
    volume = "LATTICE2024",
    pages = "415",
    year = "2025"
}

@article{Bulgarelli:2024brv,
    author = "Bulgarelli, Andrea and Cellini, Elia and Nada, Alessandro",
    title = "{Scaling of stochastic normalizing flows in SU(3) lattice gauge theory}",
    eprint = "2412.00200",
    archivePrefix = "arXiv",
    primaryClass = "hep-lat",
    doi = "10.1103/PhysRevD.111.074517",
    journal = "Phys. Rev. D",
    volume = "111",
    number = "7",
    pages = "074517",
    year = "2025"
}

@article{Bonanno:2025pdp,
    author = "Bonanno, Claudio and Bulgarelli, Andrea and Cellini, Elia and Nada, Alessandro and Panfalone, Dario and Vadacchino, Davide and Verzichelli, Lorenzo",
    title = "{Scaling flow-based approaches for topology sampling in SU(3) gauge theory}",
    eprint = "2510.25704",
    archivePrefix = "arXiv",
    primaryClass = "hep-lat",
    doi = "10.1007/JHEP04(2026)051",
    journal = "JHEP",
    volume = "04",
    pages = "051",
    year = "2026"
}

@article{PhysRevLett.78.2690,
  title = {Nonequilibrium Equality for Free Energy Differences},
  author = {Jarzynski, C.},
  journal = {Phys. Rev. Lett.},
  volume = {78},
  issue = {14},
  pages = {2690--2693},
  numpages = {0},
  year = {1997},
  month = {Apr},
  publisher = {American Physical Society},
  doi = {10.1103/PhysRevLett.78.2690},
  url = {https://link.aps.org/doi/10.1103/PhysRevLett.78.2690},
  eprint = "cond-mat/9610209",
  archivePrefix = "arXiv",
  primaryClass = "cond-mat"
}

@article{Caselle:2016wsw,
    author = "Caselle, Michele and Costagliola, Gianluca and Nada, Alessandro and Panero, Marco and Toniato, Arianna",
    title = "{Jarzynski{\textquoteright}s theorem for lattice gauge theory}",
    eprint = "1604.05544",
    archivePrefix = "arXiv",
    primaryClass = "hep-lat",
    reportNumber = "CP3-ORIGINS-2016-020, DIAS-2016-20",
    doi = "10.1103/PhysRevD.94.034503",
    journal = "Phys. Rev. D",
    volume = "94",
    number = "3",
    pages = "034503",
    year = "2016"
}

@article{Caselle:2022acb,
    author = "Caselle, Michele and Cellini, Elia and Nada, Alessandro and Panero, Marco",
    title = "{Stochastic normalizing flows as non-equilibrium transformations}",
    eprint = "2201.08862",
    archivePrefix = "arXiv",
    primaryClass = "hep-lat",
    doi = "10.1007/JHEP07(2022)015",
    journal = "JHEP",
    volume = "07",
    pages = "015",
    year = "2022"
}

@article{Holland:2025fsa,
    author = {Holland, Kieran and Ipp, Andreas and M{\"u}ller, David I. and Wenger, Urs},
    title = "{Machine-Learned Renormalization-Group-Improved Gauge Actions and Classically Perfect Gradient Flows}",
    eprint = "2504.15870",
    archivePrefix = "arXiv",
    primaryClass = "hep-lat",
    doi = "10.1103/k41k-2pnc",
    journal = "Phys. Rev. Lett.",
    volume = "136",
    number = "3",
    pages = "031901",
    year = "2026"
}

@article{Wenger:2025sre,
    author = {Wenger, Urs and Holland, Kieran and Ipp, Andreas and M{\"u}ller, David I.},
    title = "{HMC and gradient flow with machine-learned classically perfect fixed point actions}",
    eprint = "2502.03315",
    archivePrefix = "arXiv",
    primaryClass = "hep-lat",
    doi = "10.22323/1.466.0466",
    journal = "PoS",
    volume = "LATTICE2024",
    pages = "466",
    year = "2025"
}

@article{Holland:2024muu,
    author = {Holland, Kieran and Ipp, Andreas and M{\"u}ller, David I. and Wenger, Urs},
    title = "{Machine learning a fixed point action for SU(3) gauge theory with a gauge equivariant convolutional neural network}",
    eprint = "2401.06481",
    archivePrefix = "arXiv",
    primaryClass = "hep-lat",
    doi = "10.1103/PhysRevD.110.074502",
    journal = "Phys. Rev. D",
    volume = "110",
    number = "7",
    pages = "074502",
    year = "2024"
}

@article{Holland:2023ews,
    author = {Holland, Kieran and Ipp, Andreas and M{\"u}ller, David I. and Wenger, Urs},
    title = "{Fixed point actions from convolutional neural networks}",
    eprint = "2311.17816",
    archivePrefix = "arXiv",
    primaryClass = "hep-lat",
    doi = "10.22323/1.453.0038",
    journal = "PoS",
    volume = "LATTICE2023",
    pages = "038",
    year = "2024"
}

@article{Favoni:2020reg,
    author = {Favoni, Matteo and Ipp, Andreas and M{\"u}ller, David I. and Schuh, Daniel},
    title = "{Lattice Gauge Equivariant Convolutional Neural Networks}",
    eprint = "2012.12901",
    archivePrefix = "arXiv",
    primaryClass = "hep-lat",
    doi = "10.1103/PhysRevLett.128.032003",
    journal = "Phys. Rev. Lett.",
    volume = "128",
    number = "3",
    pages = "032003",
    year = "2022"
}

@article{Hasenfratz:1993sp,
    author = "Hasenfratz, P. and Niedermayer, F.",
    title = "{Perfect lattice action for asymptotically free theories}",
    eprint = "hep-lat/9308004",
    archivePrefix = "arXiv",
    reportNumber = "BUTP-93-17",
    doi = "10.1016/0550-3213(94)90261-5",
    journal = "Nucl. Phys. B",
    volume = "414",
    pages = "785--814",
    year = "1994"
}

@article{DeGrand:1995ji,
    author = "DeGrand, Thomas A. and Hasenfratz, Anna and Hasenfratz, Peter and Niedermayer, Ferenc",
    title = "{The Classically perfect fixed point action for SU(3) gauge theory}",
    eprint = "hep-lat/9506030",
    archivePrefix = "arXiv",
    reportNumber = "BUTP-95-14, COLO-HEP-361",
    doi = "10.1016/0550-3213(95)00458-5",
    journal = "Nucl. Phys. B",
    volume = "454",
    pages = "587--614",
    year = "1995"
}

@article{DeGrand:1995jk,
    author = "DeGrand, Thomas A. and Hasenfratz, Anna and Hasenfratz, Peter and Niedermayer, Ferenc",
    title = "{Nonperturbative tests of the fixed point action for SU(3) gauge theory}",
    eprint = "hep-lat/9506031",
    archivePrefix = "arXiv",
    reportNumber = "BUTP-95-15, COLO-HEP-362",
    doi = "10.1016/0550-3213(95)00459-6",
    journal = "Nucl. Phys. B",
    volume = "454",
    pages = "615--637",
    year = "1995"
}

@article{Blatter:1995ik,
    author = "Blatter, Marc and Burkhalter, Rudolf and Hasenfratz, Peter and Niedermayer, Ferenc",
    title = "{Instantons and the fixed point topological charge in the two-dimensional O(3) sigma model}",
    eprint = "hep-lat/9508028",
    archivePrefix = "arXiv",
    reportNumber = "BUTP-95-17",
    doi = "10.1103/PhysRevD.53.923",
    journal = "Phys. Rev. D",
    volume = "53",
    pages = "923--932",
    year = "1996"
}

@article{DeGrand:1995ab,
    author = "DeGrand, Thomas A. and Hasenfratz, Anna and Hasenfratz, Peter and Niedermayer, Ferenc",
    title = "{Fixed point actions for SU(3) gauge theory}",
    eprint = "hep-lat/9508024",
    archivePrefix = "arXiv",
    reportNumber = "BUTP-95-30, COLO-HEP-364",
    doi = "10.1016/0370-2693(95)01233-8",
    journal = "Phys. Lett. B",
    volume = "365",
    pages = "233--238",
    year = "1996"
}

@article{Blatter:1996np,
    author = "Blatter, Marc and Niedermayer, Ferenc",
    title = "{New fixed point action for SU(3) lattice gauge theory}",
    eprint = "hep-lat/9605017",
    archivePrefix = "arXiv",
    reportNumber = "BUTP-96-12",
    doi = "10.1016/S0550-3213(96)00523-8",
    journal = "Nucl. Phys. B",
    volume = "482",
    pages = "286--304",
    year = "1996"
}

@article{Niedermayer:2000yx,
    author = "Niedermayer, Ferenc and Rufenacht, Philipp and Wenger, Urs",
    title = "{Fixed point gauge actions with fat links: Scaling and glueballs}",
    eprint = "hep-lat/0007007",
    archivePrefix = "arXiv",
    reportNumber = "BUTP-2000-16",
    doi = "10.1016/S0550-3213(00)00731-8",
    journal = "Nucl. Phys. B",
    volume = "597",
    pages = "413--450",
    year = "2001"
}

@article{Ramos:2015baa,
    author = "Ramos, A. and Sint, S.",
    title = "{Symanzik improvement of the gradient flow in lattice gauge theories}",
    eprint = "1508.05552",
    archivePrefix = "arXiv",
    primaryClass = "hep-lat",
    reportNumber = "CERN-PH-TH-2015-199, TCDMATH-15--06",
    doi = "10.1140/epjc/s10052-015-3831-9",
    journal = "Eur. Phys. J. C",
    volume = "76",
    number = "1",
    pages = "15",
    year = "2016"
}

@article{Luscher:2010iy,
    author = {L\"uscher, Martin},
    title = "{Properties and uses of the Wilson flow in lattice QCD}",
    eprint = "1006.4518",
    archivePrefix = "arXiv",
    primaryClass = "hep-lat",
    reportNumber = "CERN-PH-TH-2010-143",
    doi = "10.1007/JHEP08(2010)071",
    journal = "JHEP",
    volume = "08",
    pages = "071",
    year = "2010",
    note = "[Erratum: JHEP 03, 092 (2014)]"
}

@article{BMW:2012hcm,
    author = {Bors\'anyi, Szabolcs and D\"urr, Stephan and Fodor, Zolt\'an and Hoelbling, Christian and Katz, S\'andor D. and Krieg, Stefan and Kurth, Thorsten and Lellouch, Laurent and Lippert, Thomas and McNeile, Craig},
    collaboration = "BMW",
    title = "{High-precision scale setting in lattice QCD}",
    eprint = "1203.4469",
    archivePrefix = "arXiv",
    primaryClass = "hep-lat",
    reportNumber = "ITP-BUDAPEST-657, CPT-P004-2012, WUB-12-02",
    doi = "10.1007/JHEP09(2012)010",
    journal = "JHEP",
    volume = "09",
    pages = "010",
    year = "2012"
}

@article{Spriggs:2025sea,
    author = "Spriggs, Thomas and Greplova, Eliska and Carrasquilla, Juan and Nys, Jannes",
    title = "{Accurate Ground States of SU(2) Lattice Gauge Theory in 2+1D and 3+1D}",
    eprint = "2509.12323",
    archivePrefix = "arXiv",
    primaryClass = "hep-lat",
    doi = "10.1103/wsk2-qcvy",
    journal = "Phys. Rev. Lett.",
    volume = "136",
    number = "10",
    pages = "101902",
    year = "2026"
}

@misc{lou2023scalingriemanniandiffusionmodels,
      title={Scaling Riemannian Diffusion Models}, 
      author={Aaron Lou and Minkai Xu and Stefano Ermon},
      year={2023},
      eprint={2310.20030},
      archivePrefix={arXiv},
      primaryClass={cs.LG},
      url={https://arxiv.org/abs/2310.20030}, 
}

@article{Hasenbusch:2017unr,
    author = "Hasenbusch, Martin",
    title = "{Fighting topological freezing in the two-dimensional CPN-1 model}",
    eprint = "1706.04443",
    archivePrefix = "arXiv",
    primaryClass = "hep-lat",
    doi = "10.1103/PhysRevD.96.054504",
    journal = "Phys. Rev. D",
    volume = "96",
    number = "5",
    pages = "054504",
    year = "2017"
}

@article{Abbott:2022zhs,
    author = "Abbott, Ryan and others",
    title = "{Gauge-equivariant flow models for sampling in lattice field theories with pseudofermions}",
    eprint = "2207.08945",
    archivePrefix = "arXiv",
    primaryClass = "hep-lat",
    reportNumber = "MIT-CTP/5446, INT-PUB-22-017",
    doi = "10.1103/PhysRevD.106.074506",
    journal = "Phys. Rev. D",
    volume = "106",
    number = "7",
    pages = "074506",
    year = "2022"
}

@article{Abbott:2022zsh,
    author = "Abbott, Ryan and others",
    title = "{Aspects of scaling and scalability for flow-based sampling of lattice QCD}",
    eprint = "2211.07541",
    archivePrefix = "arXiv",
    primaryClass = "hep-lat",
    reportNumber = "MIT-CTP/5496",
    doi = "10.1140/epja/s10050-023-01154-w",
    journal = "Eur. Phys. J. A",
    volume = "59",
    number = "11",
    pages = "257",
    year = "2023"
}

\end{document}